\documentclass[%
 reprint,
 amsmath,amssymb,
 aps,
]{revtex4-2}
\usepackage{color}
\usepackage{graphicx}% Include figure files
\usepackage{dcolumn}% Align table columns on decimal point
\usepackage{braket}
\usepackage{bm}% bold math
\usepackage{qcircuit}
\usepackage{hyperref}% add hypertext capabilities
\usepackage{bbold}
\begin{document}

\preprint{APS/123-QED}

\title{Variationally Compressing Quantum Circuits to Approximate Nonadiabatic Molecular Quantum Dynamics}

\author{Joshua M. Courtney}
\email{Joshua.Courtney1@uga.edu}
\author{P. C. Stancil}

 \affiliation{
  Department of Physics and Astronomy and the Center for Simulational Physics, University of Georgia, Athens, GA 30602, USA}

\date{\today}

\begin{abstract}
Quantum simulation has begun to penetrate the field of quantum chemistry in hopes of efficiently calculating ground state energies and approximating real-time evolution. With modern research highlighting nonadiabatic dynamics, tunably approximating deep circuits representing potential landscapes becomes crucial for simulating real quantum systems. Variationally approximating unitaries allows for shallower circuits and accuracy tunable to hardware fidelity, so long as the observable quantities are preserved. We show the variational compression of Trotter terms preserve reaction rate coefficients via classical emulation of a hybrid quantum-classical optimization method, as well as fast-forwarded adiabatic dynamics on quantum hardware. 
Compressed circuits can be incorporated with product-formula-based time evolution to approximate dynamics of a particle in two coupled harmonic potentials, allowing tunability when removing high-cost qubit interactions. 
Approximate rate coefficients are recovered after substituting terms in a nonadiabatic dynamic process, giving proof-of-principle for observable preservation under variational optimization.
Attention is paid to minimizing qubit and gate-count resources.
\end{abstract}

\maketitle

\section{\label{sec:Intro} Introduction} 

Pursuit of efficient Hamiltonian simulation has extended into numerous physical, biological, and financial fields, with vast optimizations being well-characterized for sparse, adiabatic systems \cite{low2017quantum, low2019hamiltonian, motlagh2024generalized, harrow2009quantum, berry2007efficient, berry2024doubling, zhao2022hamiltonian, feynman1982simulating, preskill2018quantum, stamatopoulos2020option, baiardi2023quantum}. 
Classical approaches to nonadiabatic dynamics offer approximation methods for high-dimensional state spaces and potential surfaces with non-constant curvature, including multiconfigurational time-dependent Hartree (MCTDH) and time-dependent density matrix renormalization group (TD-DMRG) \cite{tavernelli2015nonadiabatic, nelson2020non, burghardt2008multimode, lengsfield1992nonadiabatic, miller1970classical, sadri2014rovibrational, meyer2009multidimensional, molignini2025many, ren2022time, baiardi2021electron}.
These methods extend deep into multi-channel and multidimensional surfaces of quantum chemistry problems, but quickly encounter computational limits imposed by exponential resource scaling \cite{wang2015multilayer}.
In the limit of strong correlations, classical methods face the ``curse of dimensionality,'' where nonadiabatic dynamics produces highly entangled, inseparable quantum states.
Current proposals for quantum algorithms to handle nonadiabatic dynamics in first quantization suggest high hardware overhead \cite{ollitrault2020nonadiabatic, navickas2025experimental}, motivating the development of variational approaches compatible with both noisy intermediate-scale quantum (NISQ) and early fault-tolerant architectures.

Variational methods are attractive for near-term quantum computing, with ground state energy calculations pursued via variational quantum eigensolver (VQE) techniques in both circuit and measurement-based implementations \cite{wang2019accelerated, peruzzo2014variational, ferguson2021measurement, cerezo2022variational, park2024efficient, tamiya2021calculating}. 
Researchers have characterized cost landscapes and developed dressed quantum circuits using hybrid quantum-classical techniques, finding applications in phase estimation and dynamics simulation \cite{mcclean2018barren, zhang2022variational, endo2021hybrid, mcclean2016theory, endo2020variational, yao2021adaptive, benedetti2021hardware, heya2019subspace, barison2021efficient}. 
Quantum computational chemistry has increasingly adopted variational algorithms, with research into nonadiabatic systems connecting to other dynamical problems such as quantum scattering \cite{mcardle2020quantum, cao2019quantum, liu2022prospects, parrish2019hybrid, cook2016towards, o2019calculating, navickas2025experimental, lengsfield1992nonadiabatic}. 

Several encoding strategies exist for representing quantum dynamics on qubit registers.
Some methods optimize for storing a particle's excitations, while others represent wavepacket propagation through probability density evolution on a binary-encoded grid~\cite{stamatopoulos2020option, bornman2023introduction, harrow2009quantum, lloyd1996universal, staelens2021scattering}.
Ollitrault et al. \cite{ollitrault2020nonadiabatic} demonstrated nonadiabatic dynamics using binary encoding, achieving highly accurate state populations with polynomial circuit complexity.
This approach offers logarithmic qubit scaling for position storage and straightforward extension to higher-dimensional spaces, with numerical error controlled primarily by Trotter step size and wavepacket spatial resolution.  
These factors are inextricably connected through commutator norm calculation~\cite{childs2021theory}, where increasing the number of qubits representing a coordinate decreases discretization error (via the Nyquist-Shannon theorem) but increases Trotter error. 
A critical limitation remains in the explicit circuit construction for general potential surfaces, scaling as $\mathcal{O}(n^2)$ for quadratic functions with all-to-all qubit connectivity requirements that are costly on hardware with limited native interactions.
Overcoming this limitation may enable extension to other fields that benefit from wavepacket-resolved information, including high-energy physics and quantum chemical scattering \cite{pritchett2010quantum}.  

A method to tunably approximate diagonal operators would address this limitation by matching logical approximation error to physical gate error while respecting hardware connectivity.
Control over circuit structure enables mapping to native gate sets with minimal compilation overhead and improved compatibility with error mitigation techniques. 
By producing shallower circuits with reduced complexity, variational compression offers practical value for near-term quantum simulation of chemically relevant dynamic systems, an important stop on the road to quantum advantage alongside the static observables sought by VQE.

In this article, we develop classical and quantum-classical hybrid methods to variationally compress quadratic operators applied to binary-encoded qubit registers.
We employ a diagonalized ansatz circuit from variational fast-forwarding (VFF), which has found experimental verification in spectral decomposition and applies broadly to sparse Hamiltonians \cite{atia2017fast, cirstoiu2020variational, commeau2020variational, geller2021experimental, berry2007efficient, balasubramanian2020quantum}. 
We optimize this ansatz using quantum-assisted quantum compiling (QAQC), adapting the entanglement fidelity measure to calculate cost functions and gradients \cite{khatri2019quantum, sharma2020noise}.

Contributions of this work include the variational compression of Trotterized quadratic operators, prioritizing proximal connectivity to minimize SWAP overhead or routing costs. 
We demonstrate that compressed circuits are fast-forwardable, enabling time evolution for multiple steps by rescaling diagonal parameters without reoptimization, experimentally validating on $ibm\_brisbane$ for $n = 3$ qubits.
We recover Marcus model rate coefficients using truncated ansatzes through classical emulation, demonstrating preservation of the observable under controlled approximation. 
Finally, we give explicit gate counts to compare compressed circuits to explicit constructions and quantify depth reduction for NISQ implementations.

We demonstrate these methods on the Marcus model to describe vibronic state transfer dynamics commonly encountered in electron transfer reactions.
The ansatz we employ contains long-range entanglement operations that would require SWAP gates on hardware with limited connectivity.
The truncation removes these unfavorable interactions while reparameterizing the remaining gates to preserve simulation quality, enabling circuit adaptation to specific quantum hardware architectures.
Our approach further directly reparameterizes controlled-phase based gates to parity-based gates, highlighting a proof-of-principle for QAQC to optimize for native gate sets. 

Building upon existing methods (VFF \cite{cirstoiu2020variational}, QAQC \cite{khatri2019quantum}, Walsh operator expansions \cite{walsh1923closed}, and the Marcus model benchmark \cite{ollitrault2020nonadiabatic}), we develop an $l$-local truncation strategy for diagonal Walsh operators on binary-encoded registers, enabling tunable removal of long-range entangling gates. 
We show that quadratic operators on binary-encoded grids reduce to two effective variational parameters under truncation, independent of register size, and that the substitution of compressed operators into nonadiabatic Trotter circuits retain observable benchmarks via Marcus model rate coefficients. 
Apart from the approximate nonadiabatic dynamics, we show that free-particle propagation is fast-forwardable on IBM quantum hardware, demonstrating noise resilience of the ansatz structure.

The remainder of this paper is organized as follows.
A brief background of binary encoding of a grid and Hamiltonian decomposition of the Marcus model is given in Section~\ref{sec:Background}.
Section~\ref{sec:Method} details compression methodology, including ansatz construction, truncation strategy, QAQC optimization, and hardware implementations.
Section~\ref{sec:Results} presents numerical results for circuit optimization and classical emulation, along with experimental results from IBM quantum hardware.
We analyze resource scaling and limitations in Section~\ref{sec:Discussion}, discussing the implications for near-term quantum simulation, with a summary and path for future work in Section~\ref{sec:Conclusions}.

\section{\label{sec:Background} Background}
\paragraph{Binary Encoding of Position Space --}

A one-dimensional wavepacket on a register of $n$ qubits can be represented using binary encoding.
Each computational basis state $\ket{k_{n-1} \ldots k_1 k_0}$ with $k_j \in \{0, 1\}$ corresponds to a discrete position
\begin{equation}
    x = \Delta \sum_{j=0}^{n-1} 2^j k_j,
    \label{eq:binary_position}
\end{equation}
where $\Delta = L/2^n$ is the grid spacing and $L$ is the total box size. This encoding maps $2^n$ discrete positions onto $n$ qubits, providing exponential compression of the classical grid representation.

The momentum-space representation follows from the discrete Fourier transform. Applying a centered quantum Fourier transform (cQFT) to the position register yields momentum eigenstates with the Brillouin zone centered at zero momentum, avoiding aliasing artifacts during dynamics \cite{benenti2008quantum}. The momentum grid spacing is $\Delta_p = 2\pi/L$, with Nyquist limits $\pm \pi/\Delta$ determining the maximum resolvable momentum.

\paragraph{Trotterization of Quantum Dynamics --}

We simulate time evolution under a Hamiltonian $H$ using product formula methods. For a Hamiltonian decomposed as $H = \sum_k H_k$, the first-order Trotter-Suzuki formula approximates the time evolution operator as
\begin{equation}
    e^{-iH\tau} \approx \prod_{k} e^{-iH_k \tau} + \mathcal{O}(\tau^2),
    \label{eq:trotter}
\end{equation}
where $\tau$ is the timestep. For kinetic and potential terms, this becomes $e^{-iT\tau}e^{-iV\tau} + \mathcal{O}(\tau^2)$. Higher-order formulas achieve better accuracy at the cost of additional circuit depth \cite{lloyd1996universal}.

\begin{figure}[t]
\includegraphics[width=\columnwidth]{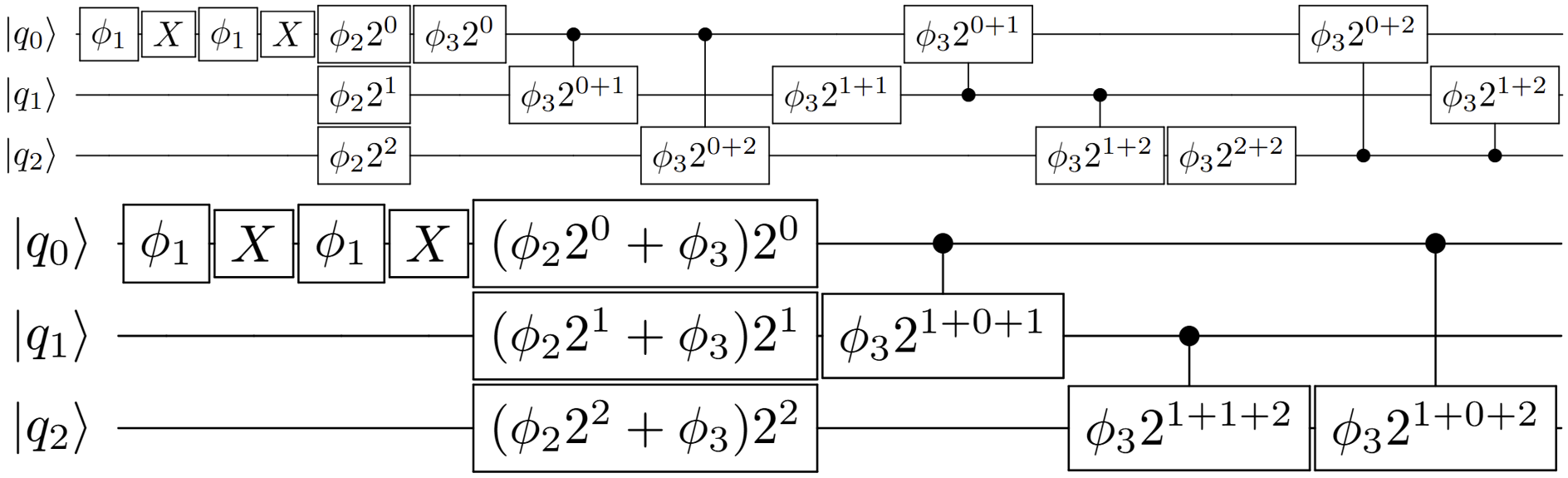}
\caption{\label{fig:explicit_quadratic} 
Explicit circuit decomposition for the same quadratic operator on $n=3$ qubits. 
Operations are ordered by polynomial degree: global phase ($\phi_1$), linear terms ($\phi_2$), and quadratic terms ($\phi_3$). 
The quadratic terms require all-to-all connectivity between qubits.
(above) Quadratic circuit as presented by Benenti and Strini~\cite{benenti2008quantum} and Ollitrault et al.~\cite{ollitrault2020nonadiabatic}.
(below) Tensor product commutativity reduces resources with an equivalently expressive circuit.}
\end{figure}

For nonadiabatic dynamics with kinetic operator $K$ and potential operators $V_i$, each Trotter term $e^{-iH_k \tau}$ must be implemented as a quantum circuit. The explicit circuit construction for polynomial operators on binary-encoded registers was developed by Benenti and Strini \cite{benenti2008quantum}. A quadratic operator $f(x) = ax^2 + bx + c$ applied to the position basis yields
\begin{equation}
    e^{-i\tau f(x)} = e^{-i\tau a\Delta^2 \left(\sum_j 2^j k_j\right)^2}  \cdot e^{-i\tau b\Delta \sum_j 2^j k_j} \cdot e^{-i\tau c} .
    \label{eq:quadratic_trotter}
\end{equation}
The linear term produces single-qubit $R_z$ rotations with angles weighted by powers of 2. 
The quadratic term produces both single-qubit rotations (from qubit index idempotency $k_j^2 = k_j$) and two-qubit controlled rotations 
$k_j k_\ell$, resulting in all-to-all connectivity and $\mathcal{O}(n^2)$ circuit depth (Figure \ref{fig:explicit_quadratic}).
We immediately see through tensor product commutativity that many of these operations commute, reducing gate counts reported by Benenti and Strini \cite{benenti2008quantum} and Ollitrault et al. \cite{ollitrault2020nonadiabatic} from $2(1+n)$ $R_z$ gates and $n(n-1)$ $CR_z$ gates to $2+n$ $R_z$ gates and $n(n-1)/2$ $CR_z$ gates.
This gives a large constant factor of analytic compression prior to variational optimization.

\paragraph{Variational Ansatz Construction --}
For matrices that are both unitary and Hermitian, the Schur decomposition is a classic diagonalization $U = WDW^\dagger$,
where $W$ is unitary and $D$ is diagonal containing the eigenvalues of $U$. 
This decomposition is unique up to eigenvalue ordering and eigenvector phases, motivating the variational ansatz structure.
By tuning parameters for both the eigenvector matrix $W$ and the diagonal matrix $D$ for a target unitary $U$, we obtain a compressed representation that separates the ``rotation into the eigenbasis'' (via $W$) from the ``phase accumulation'' (via $D$). 
Time evolution becomes trivial in the diagonal representation, where evolving for time $N\tau$ rescales the diagonal phases by $N$, enabling fast-forwarding \cite{atia2017fast}.

We decide ansatzes for $W$ and $D$ using Walsh functions, which form a complete orthonormal basis for functions on binary strings, analogous to the Fourier basis for periodic functions \cite{walsh1923closed}. 
For $n$ binary variables $k_0, k_1, \ldots, k_{n-1} \in \{0,1\}$, the Walsh functions are
\begin{equation}
    W_j(k_0, \ldots, k_{n-1}) = (-1)^{\sum_{q=0}^{n-1} j_q k_q},
    \label{eq:walsh_functions}
\end{equation}
where $j = \sum_{q=0}^{n-1} j_q 2^q$ indexes the function and $j_q \in \{0,1\}$ are its binary digits. 

Any real-valued function can be expanded in the Walsh basis:
\begin{equation}
    f(k_0, \ldots, k_{n-1}) = \sum_{j=0}^{2^n - 1} \hat{f}_j \, W_j(k_0, \ldots, k_{n-1}),
    \label{eq:walsh_expansion}
\end{equation}
with Walsh coefficients $\hat{f}_j = 2^{-n} \sum_k f(k) W_j(k)$.

For quantum circuits, this translates to tensor products of Pauli-$Z$ operators.
Since $Z\ket{k} = (-1)^k \ket{k}$ for $k \in \{0,1\}$, we have
\begin{equation}\label{eq:walsh_pauli}
\begin{split}
    &W_j(k_0, \ldots, k_{n-1})  =
    \\&\bra{k_0 \ldots k_{n-1}} \bigotimes_{q=0}^{n-1} Z_q^{j_q} \ket{k_0 \ldots k_{n-1}}.
\end{split}
\end{equation}
A diagonal unitary $e^{-i\tau f(x)}$ can therefore be written as
\begin{equation}
    e^{-i\tau f(x)} = \prod_{j=0}^{2^n-1} e^{-i\tau \hat{f}_j \bigotimes_{q=0}^{n-1} Z_q^{j_q}}.
    \label{eq:diagonal_walsh}
\end{equation}

The order of a Walsh term is defined as its Hamming weight $|h| = \sum_q h_q$, corresponding to the number of qubits involved in the Pauli-$Z$ tensor product, with order 0 being the identity, order 1 being $R_z$ rotations, and order 2 being parity-based $ZZ$ gates.
Quadratic functions on binary-encoded positions have Walsh expansions truncated at order $2$, giving a scheme for approximating diagonal unitaries.
An ansatz from a Walsh function expansion placed in a Schur decomposition gives a fast-forwardable, variationally optimizable circuit, following Cirstoiu et al. \cite{cirstoiu2020variational}.

\paragraph{Variational Fast Forwarding --}

Following a Schur decomposition, our VFF ansatz takes the form
\begin{equation}
    A(\vec{\gamma}, \vec{\theta}, \tau) = W(\vec{\gamma}) D(\vec{\theta}, \tau) W^\dagger(\vec{\gamma}),
    \label{eq:vff_ansatz}
\end{equation}
where $W(\vec{\gamma})$ is a parameterized eigenvector unitary and $D(\vec{\theta}, \tau)$ is an independently parameterized diagonal.

The eigenvector unitary $W(\vec{\gamma})$ is constructed using the Baker-Campbell-Hausdorff (BCH) formula:
\begin{equation}
    e^A e^B = e^{A + B + \frac{1}{2}[A,B] + \frac{1}{12}[A,[A,B]] - \frac{1}{12}[B,[A,B]] + \cdots}.
    \label{eq:bch}
\end{equation}
This results in a hardware-efficient ansatz (HEA) structure, using alternating layers of single-qubit gates and nearest-neighbor entangling gates to increase expressivity while respecting hardware connectivity constraints \cite{kandala2017hardware}.

The diagonal operator $D(\vec{\theta}, \tau)$ is parameterized as a truncated Walsh expansion. 
Once the ansatz is optimized for timestep $\tau$, evolution for time $N\tau$ can be achieved by simply rescaling the diagonal parameters to $N\cdot\vec{\theta}$, while $W(\vec{\gamma})$ remains unchanged. 
This enables simulation of dynamics beyond the coherence time of the quantum hardware, limited only by the accuracy of the initial optimization for compatible Hamiltonians.
To optimize these variational parameters, we look at the quantum-classical hybrid optimization algorithm of quantum-assisted quantum compiling.

QAQC provides a method to optimize variational circuits using quantum hardware to evaluate the cost function and gradients \cite{khatri2019quantum}. 
This optimization updates parameters in an ansatz $A(\vec{\alpha})$ to approximate a target unitary $U$.

The Local Hilbert-Schmidt Test (LHST) cost function measures average entanglement fidelity across qubit pairs:
\begin{equation}
    C_{\text{LHST}}(U, A) = 1 - \frac{1}{n} \sum_{j=1}^{n} F_e^{(j)},
    \label{eq:lhst_cost}
\end{equation}
where $F_e^{(j)}$ is the entanglement fidelity for the $j$-th qubit pair. 
The LHST cost satisfies $C_{\text{LHST}} = 0$ if and only if $A = e^{i\phi} U$ for some global phase $\phi$, making it a faithful measure of unitary equivalence \cite{khatri2019quantum}.

\begin{figure}[t]
\centerline{
\includegraphics[width=65mm]{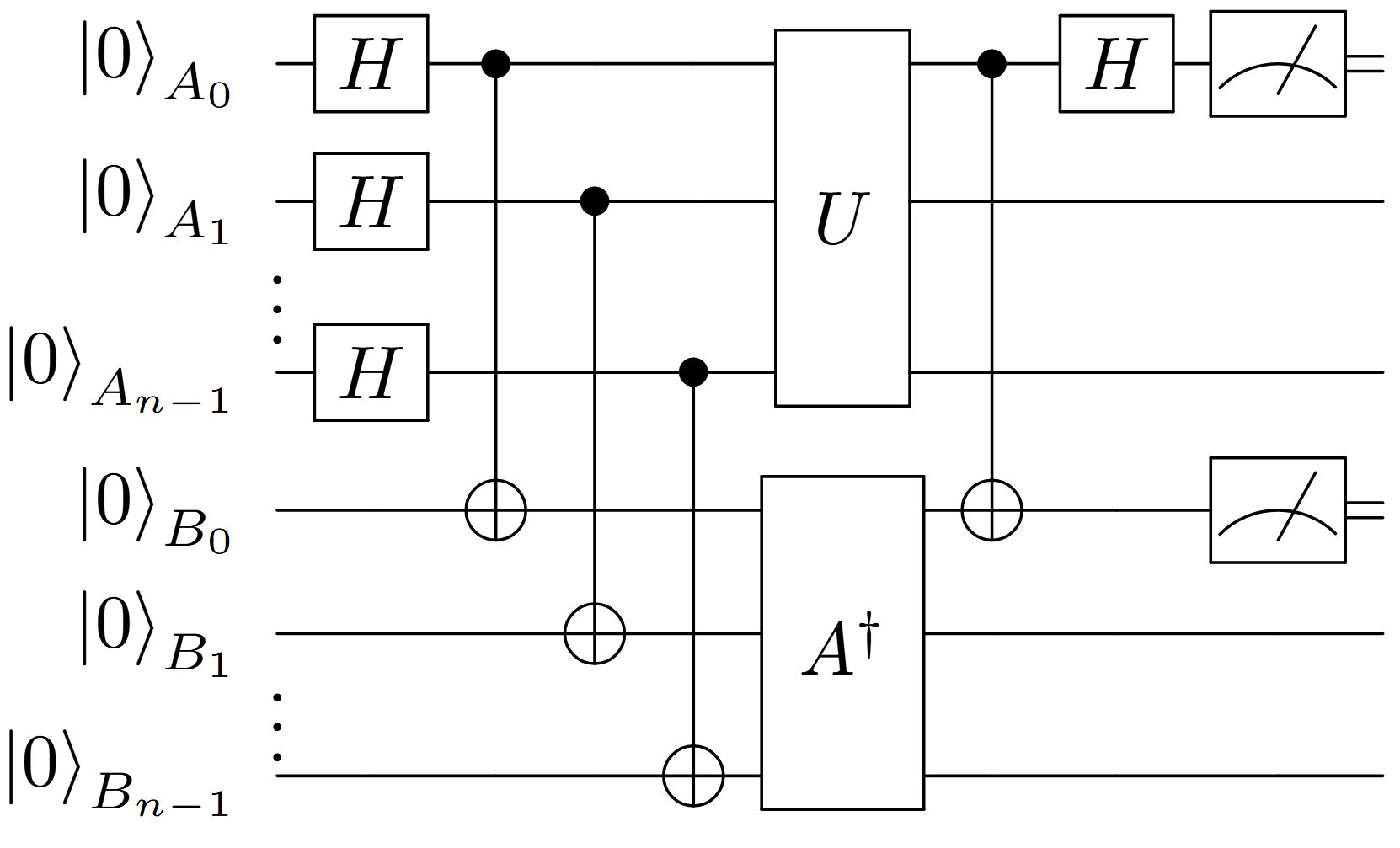}}
\caption{\label{fig:qaqc_schematic} 
Quantum-assisted quantum compiling (QAQC) circuit for measuring entanglement fidelity. 
Bell pairs are prepared between corresponding qubits in registers $A$ and $B$. 
Bell basis measurement extracts entanglement fidelity $F_e^{(j)}$ for each qubit pair, requiring measurement of all $n$ pairs of qubits.}
\end{figure}

Entanglement fidelity is measured by preparing Bell pairs between corresponding qubits in two $n$-qubit registers, applying the target $U$ to register $A$ and the ansatz inverse $A^\dagger$ to register $B$, and measuring the probability of returning to the initial Bell state (Figure~\ref{fig:qaqc_schematic}). 
This requires $2n$ qubits total and provides a noise-resilient cost function: hardware noise affects both $U$ and $A$ similarly, partially canceling in the fidelity measurement \cite{sharma2020noise}.

Gradients for optimization can be computed using the parameter-shift rule,
\begin{equation}
    \frac{\partial C}{\partial \alpha_k} = \frac{1}{2} \left[ C(\alpha_k + \pi/2) - C(\alpha_k - \pi/2) \right],
    \label{eq:parameter_shift}
\end{equation}
enabling gradient-based optimization using only cost function measurements.
We use QAQC, as a gradient descent algorithm, in a classical emulation to find the global minimum for kinetic and potential operators.

\paragraph{Marcus Model for Nonadiabatic Dynamics --}
The Marcus model describes electron transfer reactions between donor and acceptor states coupled through a reaction coordinate \cite{marcus1956theory, tachiya1993generalization}. 
In the diabatic representation, two electronic states $\ket{0}$ and $\ket{1}$ are associated with harmonic potential energy surfaces $V_0(x)$ and $V_1(x)$ along a nuclear coordinate $x$.

\begin{figure}[t]
\centerline{
\includegraphics[width=65mm]{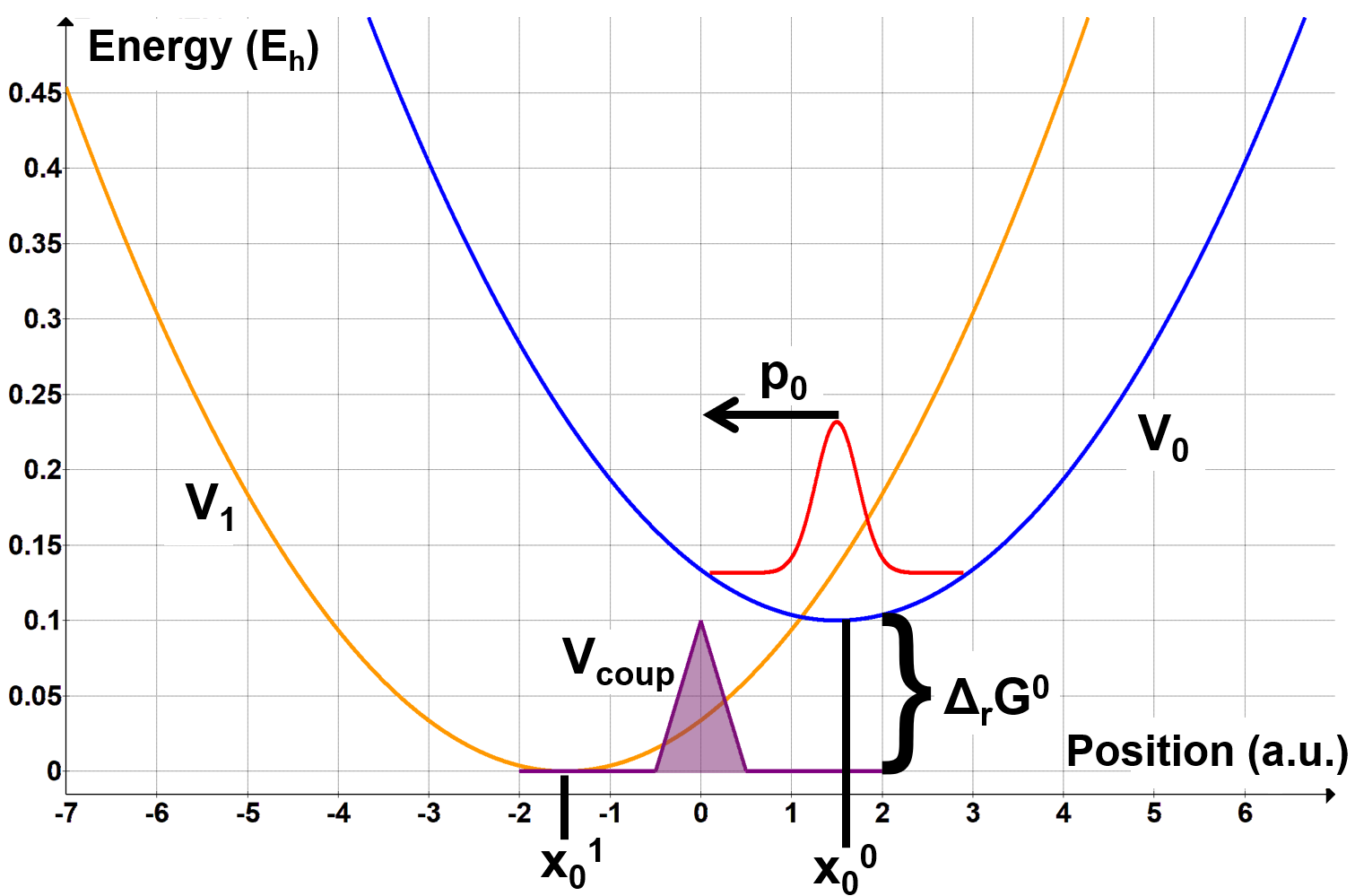}}
\caption{Marcus model for nonadiabatic dynamics with a linearly-approximated Gaussian coupling. 
Two diabatic potential energy surfaces $V_0$ and $V_1$ are offset by $\Delta_r G^{0}$.
\label{fig:marcus_model}}
\end{figure}

The diabatic potentials take the form
\begin{equation}
    V_i(x) = \frac{1}{2} \mu \omega_i^2 (x - x_i^0)^2 + \Delta_r G_i^0,
    \label{eq:diabatic_potentials}
\end{equation}
where $\mu$ is the effective mass, $\omega_i$ is the angular frequency, $x_i^0$ is the equilibrium position, and $\Delta_r G_i^0$ is the energy offset for surface $i$. The electronic coupling $V_{\text{coup}}(x)$ is typically localized near a defined reaction center or near the crossing point of the two surfaces (Figure~\ref{fig:marcus_model}).

The full Hamiltonian in the diabatic basis is
\begin{equation}
    H = K \otimes \mathbb{1}_{\text{el}} + V_0 \otimes \ket{0}\bra{0} + V_1 \otimes \ket{1}\bra{1} + V_{\text{coup}} \otimes \sigma_x,
    \label{eq:marcus_hamiltonian}
\end{equation}
where $K = p^2/2\mu$ is the kinetic energy, $\mathbb{1}_{\text{el}}$ is the identity on the electronic subspace, and $\sigma_x$ couples the diabatic states.

Quantum simulation of this system proceeds by encoding the nuclear wavepacket on a position register of $n$ qubits, using an objective qubit to store electronic state populations, implement the Trotterized evolution, then measuring the objective qubit to extract population transfer dynamics.
Population of the initial diabatic state as a function of time yields rate coefficients via short-time dynamics \cite{nitzan2024chemical}. 
Previous classical emulations of this model achieved accurate population dynamics using explicit Trotterization on 8-qubit position registers \cite{ollitrault2020nonadiabatic}, establishing a benchmark for our compressed circuit approach.

\section{\label{sec:Method} Methods}
\begin{figure}[t]
\includegraphics[width=\columnwidth]{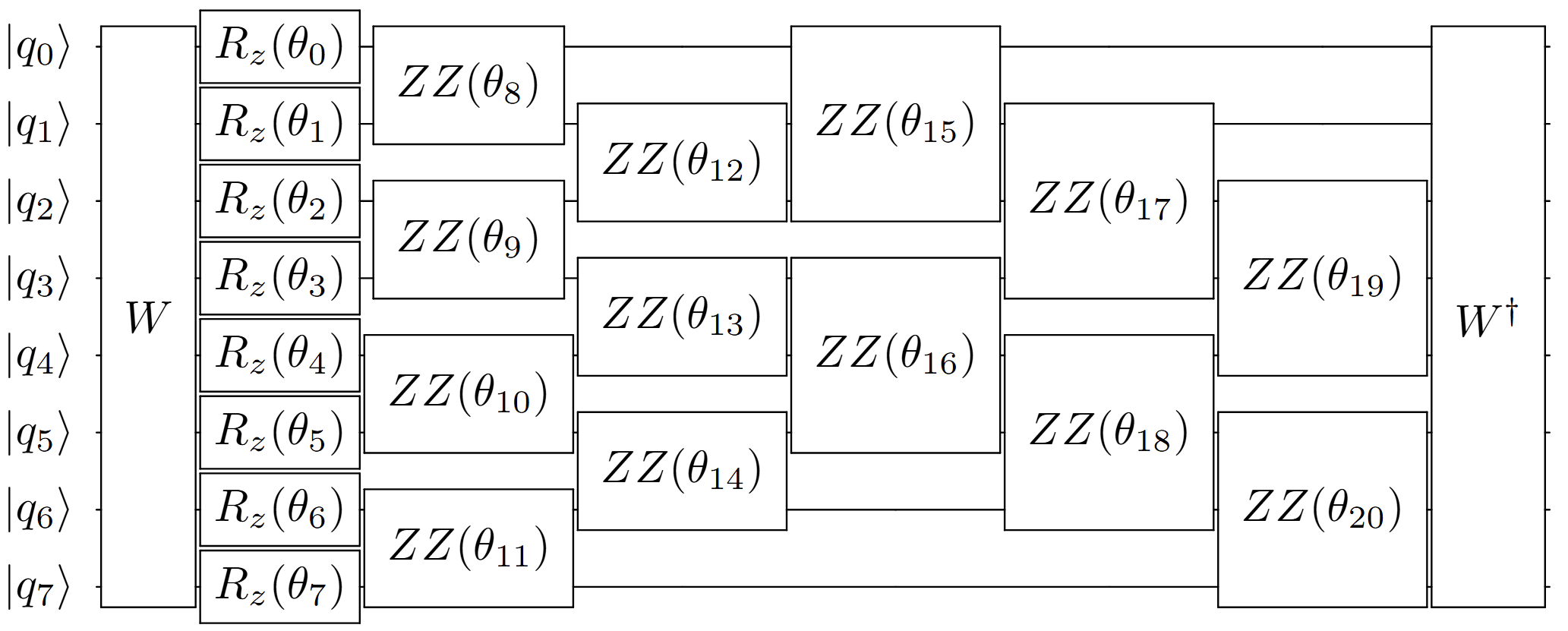}
\caption{\label{fig:l-local-ansatz} 
Ansatz structure for $l=3$-local truncation on $n=8$ qubits. 
Gates are organized by locality: $l=1$ (single-qubit $R_z$), $l=2$ (nearest-neighbor $ZZ$), and $l=3$ (next-nearest-neighbor $ZZ$).
The structure $W$ follows the first three sublayers of $D$, truncated at $l=2$-local $ZZ$ terms.}
\end{figure}

We first focus on variationally optimizing each term $T,V_0,V_1$ individually using a VFF-compatible ansatz (Figure \ref{fig:l-local-ansatz}).

Each layer of $W$ consists of three sublayers applied in sequence, being: (i) single-qubit $R_z(\gamma_i)$ rotations on each qubit $i \in \{0, 1, \ldots, n-1\}$, (ii) two-qubit $ZZ(\gamma_j)$ gates applied to \textit{even-odd} pairs: $(q_0, q_1), (q_2, q_3), (q_4, q_5), \ldots$, and (iii) two-qubit $ZZ(\gamma_k)$ gates applied to \textit{odd-even} pairs: $(q_1, q_2), (q_3, q_4), (q_5, q_6), \ldots$.
Each $ZZ(\gamma)$ gate is decomposed as $ZZ(\gamma) = \text{CNOT}(\mathbb{1} \otimes R_z(\gamma))\text{CNOT}$, which implements a parity-based rotation \cite{cirstoiu2020variational}. 
This brick-wall structure, when repeated, provides sufficient expressivity to map target unitaries to the diagonal subspace while maintaining hardware compatibility.

We parameterize the diagonal operator $D(\vec{\theta}, \tau)$ using the Walsh expansion truncated to degree $k$ (Equation \ref{eq:diagonal_walsh}).
For a quadratic function ($k=2$), this expansion contains a global phase, single qubit rotations $R_z(\theta_i)$, and $ZZ(\theta_{ij})$ gates between each pair of qubits.
From the tensor product commutation in Figure \ref{fig:explicit_quadratic}, we see a direct association with diagonal operators. 
We do not assume a one-to-one correspondence, instead using the QAQC optimization to allow the VFF ansatz to ``learn" the expanded operator, relying on the commutation result as the global minimum of this optimization.
The reliable recovery of the global minimum demonstrates that a second-order Walsh expansion exactly represents any quadratic operator on a binary-encoded basis.
When gates are removed from the ansatz, it will still find a global minimum when optimizing on the exact quadratic operator.

\paragraph{$l$-Local Truncation --}

We define the locality of a $ZZ$ operation by the distance between its control and target qubits on a linear qubit array. 
A single-qubit $R_z$ gate is 1-local, a $ZZ$ gate between neighboring qubits is 2-local, and so forth.
Following Equation \ref{eq:diagonal_walsh} and organizing by increasing locality,
\begin{equation}
    D_l(\vec{\theta}, \tau) = \prod_{m=1}^{l} \left( \prod_{\substack{j \in S_m \\ \text{locality}(j) \leq l}} e^{i\theta_j \bigotimes_{q=1}^{n}(Z_q)^{j_q}} \right).
    \label{eq:l_local}
\end{equation}
An example of an $l=3$-local ansatz circuit for $D_l$ is given in Figure~\ref{fig:l-local-ansatz}.

This truncation reduces circuit depth by eliminating long-range $ZZ$ operations that would require SWAP gates on hardware with limited connectivity. 
We show approximation tunability when these long-range operations are removed as a primary result of this paper.

When truncating to $l < n$, the remaining variational parameters ``compensate'' during optimization. 
Specifically, for a quadratic target, the first-order terms ($R_z$ sublayer) converge to a single effective parameter $\alpha_0$, and the second-order terms ($ZZ$ sublayers) converge to a second effective parameter $\alpha_1$, weighted by respective binary weights. Both relate to the original quadratic coefficients through
\begin{equation}
    \alpha_0 \propto \phi_2 = -2\tau \eta x_0 \Delta, \quad \alpha_1 \propto \phi_3 = -\tau \eta \Delta^2,
    \label{eq:parameter_relations}
\end{equation}
where $\eta = m\omega^2/2$ for a harmonic potential. 
This reduces the effective parameter space from $\mathcal{O}(n^2)$ to two (2) for a quadratic approximation.
The recovery of this relationship through uninformed gradient descent validates the method if the global minimum is reached.

\paragraph{Optimization via Quantum-Assisted Quantum Compiling --}

We optimize the ansatz parameters using the Local Hilbert-Schmidt Test (LHST) cost function~\cite{khatri2019quantum} (Eq.~\ref{eq:lhst_cost}),
where $F_e^{(j)}$ is the entanglement fidelity for qubit pair $j$. The entanglement fidelity is measured by initializing Bell pairs between corresponding qubits in two registers, applying $U$ to one register and $A^\dagger$ to the other, and measuring the overlap with the initial Bell state:
\begin{equation}
\begin{split}\label{eq:entanglement_fidelity}
    F_e^{(j)} = \text{Tr}\Big[ \ket{\Phi^+}\bra{\Phi^+}_{A_j B_j} \left(\mathcal{E}_j \otimes I_{B_j}\right)\\\left(\ket{\Phi^+}\bra{\Phi^+}_{A_j B_j}\right) \Big],
\end{split}
\end{equation}
where $\mathcal{E}_j$ is the quantum channel acting on qubit $A_j$ after tracing out all other qubits in register $A$.

We compute gradients using the parameter-shift rule. 
For eigenvector parameters $\theta_k$, we can expand Equation \ref{eq:parameter_shift} as 
\begin{equation}
\begin{split}
    \frac{\partial C_{\text{LHST}}}{\partial \gamma_k} = \frac{1}{2} \Big[ & C_{\text{LHST}}(U, W_+^k D W^\dagger) \\
    & - C_{\text{LHST}}(U, W_-^k D W^\dagger) \\
    & + C_{\text{LHST}}(U, W D (W_+^k)^\dagger) \\
    &- C_{\text{LHST}}(U, W D (W_-^k)^\dagger) \Big],
\end{split}
\label{eq:grad_theta}
\end{equation}
where $W_\pm^k$ denotes $W$ with parameter $\theta_k$ shifted by $\pm\pi/2$. For diagonal parameters $\theta_\ell$:
\begin{equation}
\begin{split}
    \frac{\partial C_{\text{LHST}}}{\partial \theta_\ell} & = \frac{1}{2} \Big[ C_{\text{LHST}}(U, W D_+^\ell W^\dagger) \\ 
    & - C_{\text{LHST}}(U, W D_-^\ell W^\dagger) \Big].
    \label{eq:grad_gamma}
    \end{split}
\end{equation}

We employ Adam optimization with these gradients, using the adaptive learning rate to escape local minima and navigate the $\mathcal{O}(n^2) \mod 2\pi$ parameter landscape. 
For register sizes $n \leq 5$, we classically emulate the full $2n$-qubit QAQC protocol.
For $n > 5$, we directly minimize the Hilbert-Schmidt distance $\|A^\dagger U\|_{\text{HS}}^2$ and rescale to match $C_{\text{LHST}}$.
In all cases, we verify that optimizations converge to global minima by comparing the optimized diagonal parameters to analytical expressions derived from the explicit quadratic expansion (Figure \ref{fig:optimization_results}).
Once truncated, we apply truncated operators $T',V_0', V_1'$ in place of Trotter terms $T,V_0,V_1$ to approximate nonadiabatic molecular dynamics.

\begin{figure*}[t]
\includegraphics[width=\textwidth]{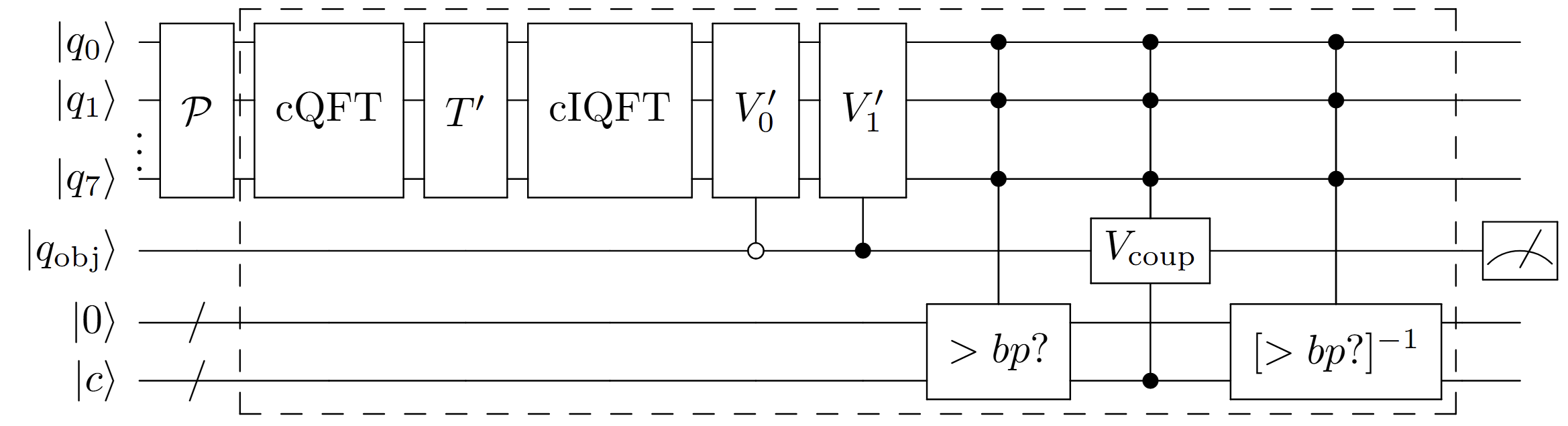}
\caption{\label{fig:trotter_circuit} 
First-order Trotter step circuit for nonadiabatic dynamics. 
The position register ($n$ qubits) encodes the wavepacket; the objective qubit stores diabatic state populations; ancilla and comparator qubits support the coupling operation. 
The approximated kinetic operator $T'$ is sandwiched between centered quantum Fourier transforms (cQFT). 
Approximated potential operators $V_0'$ and $V_1'$ are conditioned on the objective qubit state.}
\end{figure*}

\paragraph{Application to Marcus Model Dynamics --}
We simulate nonadiabatic dynamics in the Schr\"odinger picture following Ref.~\cite{ollitrault2020nonadiabatic}. 
Following Equation \ref{eq:marcus_hamiltonian}, we implement time evolution via second-order Trotterization:
\begin{equation}
    U(\tau) \approx e^{-iT\tau/2} e^{-iV_\text{diag}/2}e^{-iV_\text{coup}\tau}e^{-iV_\text{diag}/2}e^{-iT\tau/2}.
    \label{eq:trotter_circuit}
\end{equation}
The kinetic operator acts in momentum space; we sandwich it between centered quantum Fourier transforms (cQFT) to maintain the Brillouin zone center at zero momentum. 
A schematic for a first-order (simplified) schematic is given in Figure \ref{fig:trotter_circuit}).

The coupling operator $V_{\text{coup}}$ implements a position-dependent rotation on the objective qubit. 
We approximate a Gaussian coupling $V_{\text{coup}}(x) = C_0 e^{-\beta(x-a)^2}$ with an area-preserving step function spanning three basis states centered at $L/2$. 
This requires a comparison circuit to determine if the wavepacket position exceeds threshold values, and Controlled rotations on the objective qubit conditioned on comparison outcomes.

The comparison circuit uses $\lceil \log_2 n \rceil$ ancilla qubits and $P-1$ comparator qubits for a $P$-piece piecewise linear approximation. 
We construct these using Toffoli gates implementing logical AND/OR operations, achieving depth scaling of $\mathcal{O}(n^{1.5})$ for the bespoke circuits detailed in Appendix \ref{app:log-size_comparator}.
Draper arithmetic is preferred for large circuits, with $\mathcal{O}(n\log n)$ gate scaling, but the Toffoli cascades require fewer resources for $n \lesssim16$ qubits, having a smaller leading coefficient.
Toffoli cascades also require three fewer ancilla, reducing total register size from $18$ logical qubits to $15$.

\paragraph{Compressed Trotter terms --}

We replace the explicit kinetic and potential operators with their variational approximations $T'$, $V_0'$, and $V_1'$.
Each compressed term uses the $l$-local truncated ansatz from Eq.~\eqref{eq:l_local}, optimized to achieve $C_{\text{LHST}} \leq 0.01$. 
For an 8-qubit position register, the kinetic term is approximated as being $l=4$-local, and the potential terms $V_0'$,$V_1'$ are truncated to be $l=6$-local for linear qubit connectivity and $l=4$-local for a ring topology.

We calculate rate coefficients from short-time population dynamics using the approximated Trotter evolution, comparing results for step sizes $\tau = 1$ a.u. and $\tau = 10$ a.u., given in Section \ref{sec:Results}.
First, we exhibit variational fast-forwarding on $n=3$-qubits for an adiabatic evolution, and wavepacket initialization on up to $n=8$ qubits.
Nonadiabatic quantum dynamics are broadly expected to be exceptionally high overhead~\cite{kassal2008polynomial}. 
While we use $n=15$ ideal logical qubits in classical emulation, gate precision and circuit depth prevent current hardware from applying meaningful quantum simulation.

\section{\label{sec:Results} Results}

\paragraph{Fast-Forwarding Demonstration --}

\begin{figure}[t]
\includegraphics[width=\columnwidth]{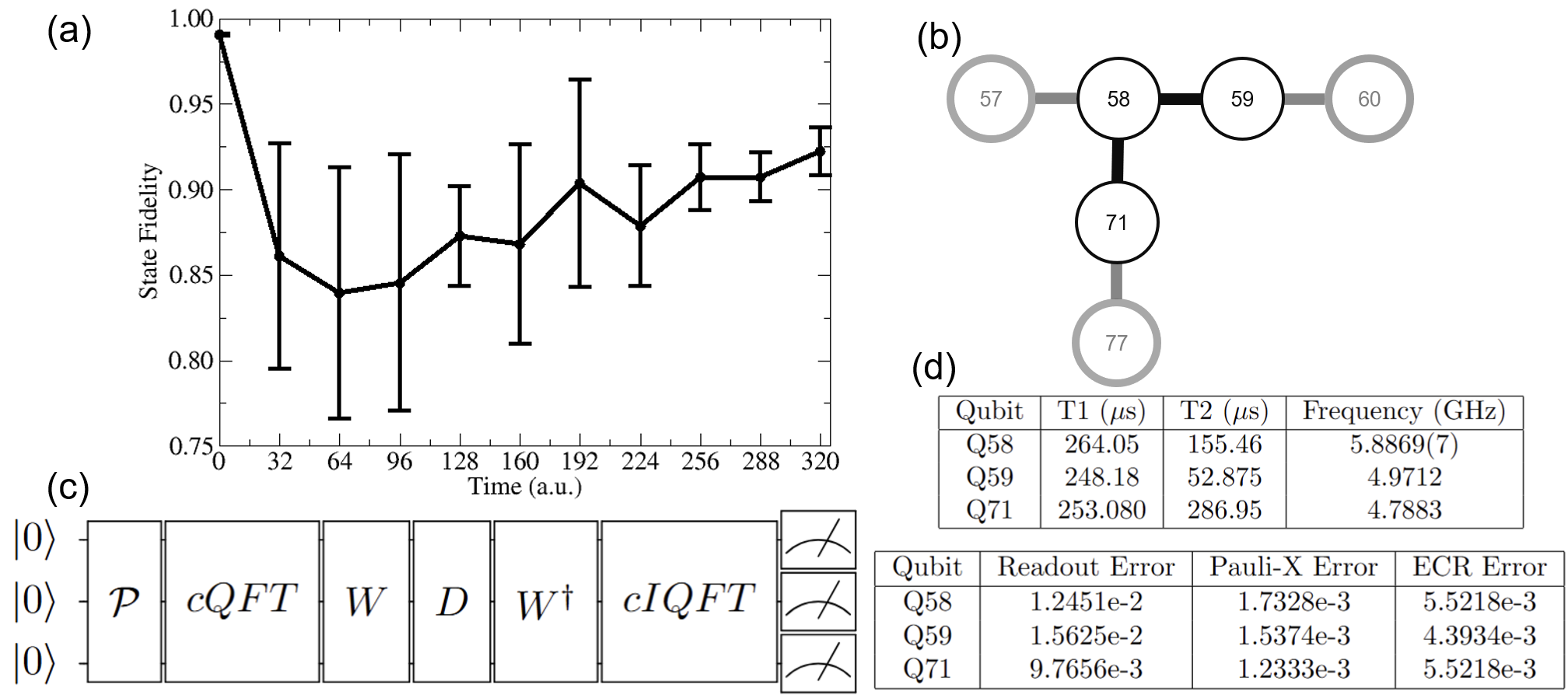}
\caption{\label{fig:fast_forwarding} 
Fast-forwarding demonstration on $ibm\_brisbane$. 
(a) State fidelity versus number of fast-forwarded timesteps, comparing hardware to noiseless simulation. 
(b) Qubit layout showing the three qubits used. 
(c) Circuit structure: UCC initialization ($\mathcal{P}$), cQFT, compressed VFF ansatz ($WDW^\dagger$), and inverse cQFT. 
(d) Calibration data for the qubits used during experiments.}
\end{figure}

We demonstrate fast-forwarding of the compressed kinetic operator on $ibm\_brisbane$, a 127-qubit Eagle processor.
Starting from the initialized wavepacket, we applied $N = 0$ to $10$ timesteps of the $l=1$-local compressed kinetic operator with $\tau = 32$ a.u., sandwiched between cQFTs (Figure \ref{fig:fast_forwarding}).

\begin{figure}[b]
\includegraphics[width=80mm]{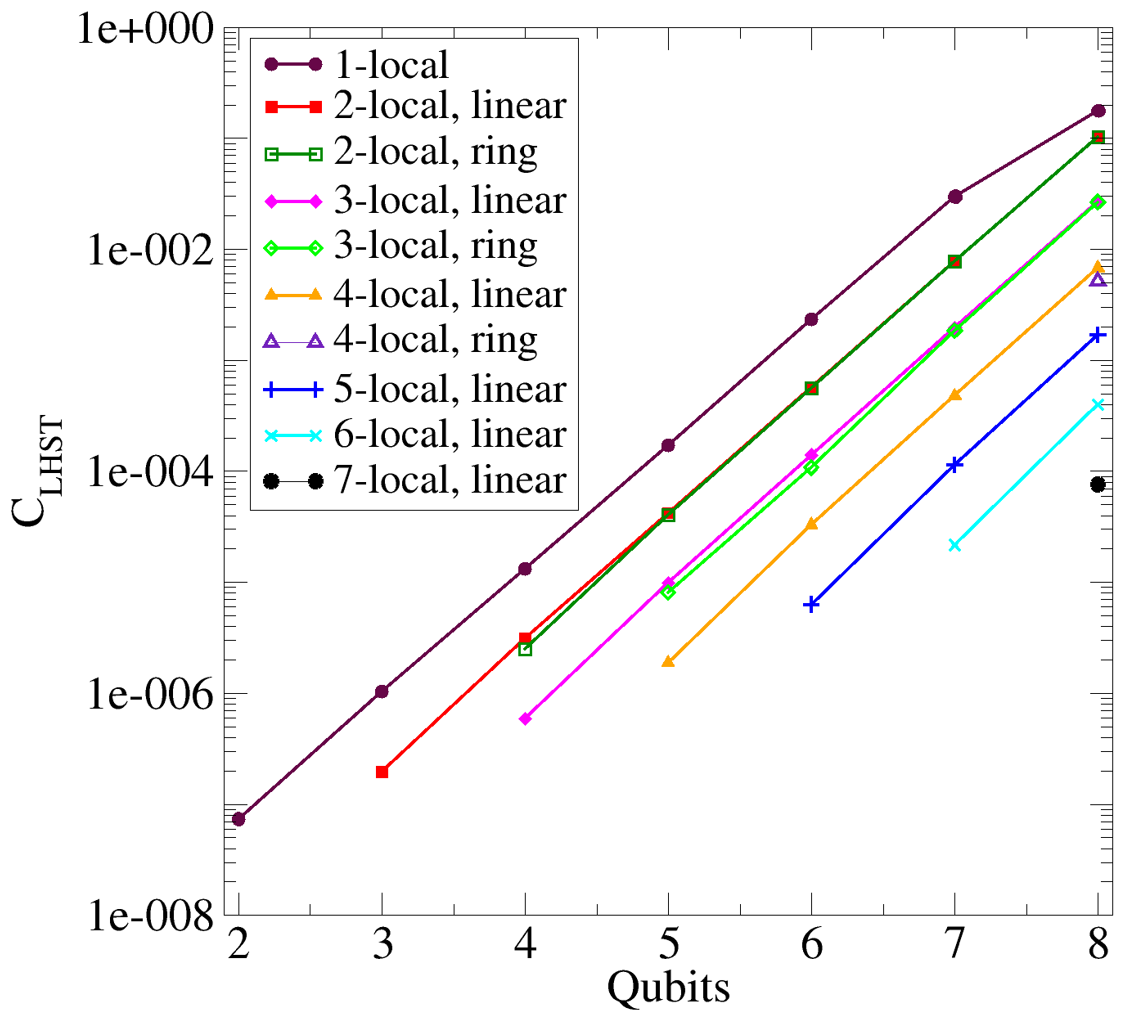}
\caption{\label{fig:optimization_results} 
LHST cost versus locality parameter $l$ for the kinetic operator. 
Costs were verified as global minima by relation of variational parameters to analytically-derived parameterizations of quadratic operators.
Solid lines indicate linear qubit topology; dashed lines indicate ring topology. 
A threshold of $C_{\text{LHST}} = 0.01$ was threshold used for compression. 
Ring topology achieves equivalent fidelity at lower $l$ due to reduced maximum qubit distance.}
\end{figure}

The state fidelity remained above 0.80 through $N = 10$ timesteps, demonstrating that the compressed ansatz structure is compatible with near-term hardware noise levels.
We emphasize that this experiment validates the noise resilience of the VFF ansatz structure itself. 
Namely, the $WDW^\dagger$ decomposition with rescaled diagonal parameters maintains fidelity under hardware noise across multiple fast-forwarded steps rather than the full nonadiabatic simulation pipeline.
A single nonadiabatic Trotter step at $n = 8$ qubits requires the compressed $T'$, $V_0'$, $V_1'$ operators, two cQFTs, the coupling circuit, and controlled operations on the objective qubit, totaling on the order of several hundred two-qubit gates (see Section~\ref{sec:Discussion}).
Running $N = 100$ such steps to extract rate coefficients would require circuit depths well beyond current coherence times on superconducting hardware, motivating continued development of error mitigation and fault-tolerant approaches alongside the circuit compression demonstrated here.

\paragraph{Optimization Convergence --}

\begin{figure}[b]
\includegraphics[width=65mm]{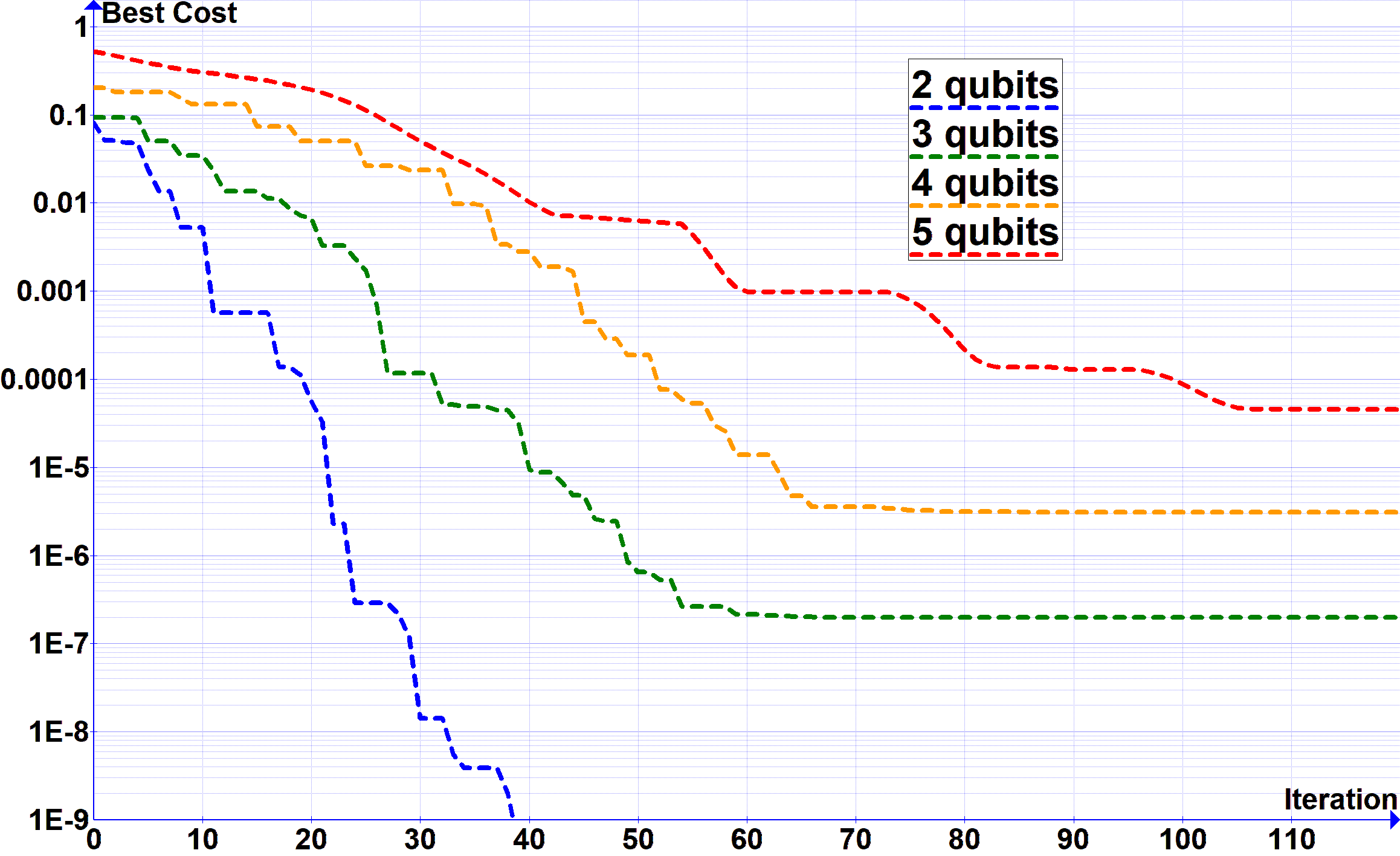}
\caption{\label{fig:qaqc_convergence} 
QAQC optimization convergence for 2-local ansatz circuits. 
The ``Best Cost'' tracks the minimum $C_{\text{LHST}}$ found at each iteration. 
Larger register sizes require more iterations due to the increased parameter space, and do not reach a zero cost from imposed connectivity constraints.}
\end{figure}

We optimize $l$-local ansatzes for the kinetic and potential operators across register sizes $n = 3$ to $n = 8$ qubits, considering both linear and ring circuit topologies. 
Figure~\ref{fig:optimization_results} shows the LHST cost as a function of locality $l$ for each configuration. 
Optimization progressions for 2-local ansatz circuits for $n=[2,5]$ are given in Figure~\ref{fig:qaqc_convergence}.

For the kinetic operator, $l = 4$-local truncation achieves $C_{\text{LHST}} < 0.01$ for all tested register sizes with linear topology. 
The ring topology achieves the same threshold at $l = 3$ for $n \leq 6$ qubits. 
The potential operators need longer-range gates for the same fidelity, requiring $l = 6$ for linear, $l = 4$ for ring topology. 
This is likely due to their spatial centering near the middle of the position basis, where binary encoding concentrates high-order bit contributions.

\paragraph{Rate Coefficient --}

We extracted Marcus model rate coefficients from the short-time population dynamics of the initial diabatic state. For small times, the population $\mathcal{P}_0(t)$ decays according to Fermi's golden rule:
\begin{equation}\label{eq:short_time}
    \mathcal{P}_0(t) \approx 1 - k t + \mathcal{O}(t^2),
\end{equation}
where $k$ is the rate coefficient. We computed $\mathcal{P}_0(t)$ by measuring the expectation value of the objective qubit, $\mathcal{P}_0 = (\langle Z_{\text{obj}} \rangle + 1)/2$, after $N$ Trotter steps of duration $\tau$.

To extract $k$, we performed linear regression on $\mathcal{P}_0(t)$ for $t \leq 100$ a.u., where the short-time approximation remains valid. The slope of the linear fit yields $-k$. We repeated this procedure for driving forces $\Delta G^0 \in [0, 0.3]$ a.u.\ to map out the rate coefficient as a function of thermodynamic driving force.

\begin{figure}[b]
\includegraphics[width=\columnwidth]{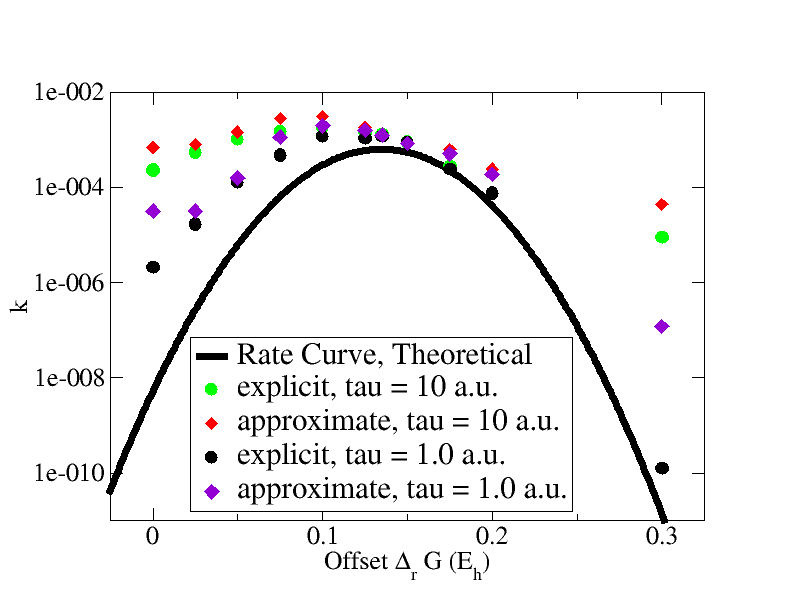}
\caption{\label{fig:rate_coefficients}
Marcus model rate coefficients extracted from short-time population dynamics.
Solid line: theoretical Marcus rate (Eq.~\eqref{eq:marcus_rate}).
Symbols: quantum simulation results using explicit circuits (blue) and compressed $l$-local ansatzes (red, purple) at different Trotter step sizes.}
\end{figure}

Figure \ref{fig:rate_coefficients} compares rate coefficients computed using explicit Trotter circuits to those obtained with compressed $l$-local ansatzes. The theoretical Marcus rate is given by~\cite{marcus1956theory, nitzan2024chemical}:
\begin{equation}
\begin{split}
    k_{\text{Marcus}} &= \frac{2\pi}{\hbar} |V_{\text{coup}}|^2 \frac{1}{\sqrt{4\pi \lambda k_B T}} \times \\
    &\exp\left(-\frac{(\Delta G^0 + \lambda)^2}{4\lambda k_B T}\right),
    \label{eq:marcus_rate}
    \end{split}
\end{equation}
where $|V_{\text{coup}}|$ is the electronic coupling matrix element, $\lambda$ is the reorganization energy, and $T$ is the temperature. For our zero-temperature wavepacket simulations, the rate is determined by the Franck-Condon overlap between the initial wavepacket and the crossing region.
Note that Eq.~\eqref{eq:marcus_rate} diverges as $T \to 0$; the theoretical curve in Figure~\ref{fig:rate_coefficients} uses the nuclear tunneling (low-temperature) limit, where the thermal activation factor $\exp[-(\Delta G^0 + \lambda)^2 / 4\lambda k_B T]$ is replaced by the squared Franck-Condon overlap integral between the vibrational ground state of $V_0$ and the vibronic states of $V_1$ near the crossing point \cite{nitzan2024chemical}.

The compressed circuits ($l=4$ for kinetic, $l=6$ for potential terms) recover rate coefficients that reproduce the qualitative dependence on $\Delta G^0$, including the peak location and the inverted-regime suppression, but with quantitative deviations from both the theoretical prediction and the explicit circuit results.
Near the peak rate ($\Delta G^0 \approx \lambda$), where the Franck-Condon overlap is largest, the compressed results most closely track the explicit Trotter values.
In the inverted regime ($\Delta G^0 > \lambda$) and the normal regime ($\Delta G^0 \ll \lambda$), deviations grow as the rate becomes more sensitive to the tails of the wavepacket distribution, where truncation-induced distortion of the potential landscape has the greatest effect.
These deviations arise from compounding two sources of error: the first-order Trotter error inherent to the product formula, and the $l$-local truncation error from removing long-range $ZZ$ interactions. Since both errors accumulate over $N$ Trotter steps, disentangling their individual contributions requires further ablation studies (e.g., evolving with exact Trotter terms but only compressed $V$, or vice versa), which we leave to future work.
The theoretical peak rate corresponds to the driving force at which the bottom of $V_0$ intersects the right side of $V_1$, where the most accurate compressed results are found.

\section{\label{sec:Discussion} Discussion}
\paragraph{Resource Scaling --}

The total qubit count for simulating nonadiabatic dynamics comprises several registers with distinct functions. A $d$-dimensional wavepacket requires $d \log_2(\mathcal{N})$ qubits to represent $\mathcal{N}$ grid points per dimension; for the one-dimensional case studied here ($d=1$), an $n$-qubit position register encodes $2^n$ discrete positions. Dynamics on $\kappa$ diabatic surfaces requires an objective register of size $\lceil \log_2(\kappa) \rceil$; for $\kappa = 2$ surfaces, a single objective qubit suffices.

The coupling operator introduces additional qubit overhead.
A piecewise linear coupling function with $P$ pieces requires $P-1$ comparator qubits to store threshold truth values, plus ancilla qubits for intermediate calculations. 
Using the bespoke comparison circuits detailed in the Supplemental Material, the ancilla register scales as $\lceil \log_2 n \rceil$ for $n > 3$ qubits.
The total qubit count for a one-dimensional, two-surface system with $P$-piece coupling is therefore
\begin{equation}
    n_{\text{total}} = n + 1 + (P-1) + \lceil \log_2 n \rceil = n + P + \lceil \log_2 n \rceil.
    \label{eq:total_qubits}
\end{equation}
For an 8-qubit position register with $\log_2M = \log_2(2) = 1$ objective qubit, 3-piece linear coupling ($P=3$), this yields $n_{\text{total}} = 8 + 1+ 3 + 3 = 15$ qubits, compared to 18 qubits reported in Ref.~\cite{ollitrault2020nonadiabatic} using a different comparator construction.

Gate count is most notably reduced in tensor product commutativity, reducing $n(n-1)$ $CR_z$ gates by a factor of two per quadratic term.
A conversion to more common $ZZ$ gates therefore requires $2n(n-1)$ CNOT gates and $n(n-1)$ $R_z$ gates.
Analytical compression reduces CNOT gate count by a factor of two, prior to SWAP gates.

Table~\ref{tab:gate_counts} summarizes the explicit and compressed gate counts for the kinetic and potential operators across register sizes $n = 4, 6, 8$, broken down by gate type.
For $n \leq l$, the compressed ansatz includes all Walsh terms and the $W/W^\dagger$ layers add overhead; compression savings appear only when $l < n$ eliminates long-range interactions requiring SWAP routing.

\begin{table}[t]
\caption{\label{tab:gate_counts} Logical gate counts for explicit (Ex.) versus $l$-local compressed (Comp.) operators on a linear qubit topology. $ZZ$ counts represent parity-rotation gates ($ZZ = \text{CNOT}\cdot R_z \cdot \text{CNOT}$). For controlled operators $V_0, V_1$, each $ZZ$ becomes a $CZZ$ requiring $2$ Toffoli gates; $W$ and $W^\dagger$ remain unconditional. The compressed ansatz uses one brick-wall layer for $W$. SWAP routing costs are architecture-dependent and not included; the maximum locality column indicates the longest-range interaction, which determines SWAP overhead on linear hardware.}
\begin{ruledtabular}
\begin{tabular}{l|l|c|c|c|c|c|c|c|c|c}
 & & & \multicolumn{2}{c}{$ZZ$/$CZZ$} & \multicolumn{2}{c}{$R_z$} & \multicolumn{2}{c}{Toffoli} & \multicolumn{2}{c}{Max $l$} \\
$n$ & Op. & $l$ & Ex. & Comp. & Ex. & Comp. & Ex. & Comp. & Ex. & Comp. \\
\hline
4 & $T$ & 4 & 12 & 6 & 16 & 10 & --- & --- & 4 & 4 \\
4 & $V_0$ & 4 & 12 & 6 & 16 & 10 & 24 & 12 & 4 & 4 \\
4 & $V_1$ & 4 & 12 & 6 & 16 & 10 & 24 & 12 & 4 & 4 \\
\hline
6 & $T$ & 4 & 24 & 12 & 30 & 18 & --- & --- & 6 & 4 \\
6 & $V_0$ & 6 & 30 & 15 & 36 & 21 & 60 & 30 & 6 & 6 \\
6 & $V_1$ & 6 & 30 & 15 & 36 & 21 & 60 & 30 & 6 & 6 \\
\hline
8 & $T$ & 4 & 36 & 18 & 44 & 26 & --- & --- & 8 & 4 \\
8 & $V_0$ & 6 & 50 & 25 & 58 & 63 & 100 & 50 & 8 & 6 \\
8 & $V_1$ & 6 & 50 & 25 & 58 & 63 & 100 & 50 & 8 & 6 \\
\end{tabular}
\end{ruledtabular}
\end{table}

Variational compression as it was performed most notably reduces SWAP gate cost.
If performed on a superconducting transmon chip, long-range interactions require $2d-3$ SWAP gates to move qubits that are a distance $d$ apart. 
For an $n=8$-qubit circuit, the $l=4$-local truncation of the kinetic term removes $4$ $l=5$-local $ZZ$ gates, $3$ $l=6$-local $ZZ$ gates, $2$ $l=7$-local $ZZ$ gates, and $1$ $l=8$-local $ZZ$ gate ($10$ gates total), saving $20$ CNOT gates and $10$ $R_z$ gates from the $ZZ$ decompositions alone, plus all SWAP routing overhead for those long-range interactions, while retaining an entanglement fidelity of $0.99$.
Controlled operations $V_0,V_1$ replace CNOT cost with Toffolis ($ZZ \rightarrow CZZ$), being far more resource-expensive.  
The $l=6$-local truncation therefore only saves $1$ $l=7$-local $CZZ$ gate, though this one operation alone requires $132$ CNOT gates and $14$ T gates for a full operation, resetting to initial positions. 

\paragraph{Comparison to Alternative Methods --}

Our variational compression approach offers distinct tradeoffs compared to other methods for future quantum simulation. 
Quantum signal processing (QSP) achieves optimal query complexity for Hamiltonian simulation but requires oracle access to block-encoded Hamiltonians \cite{low2017quantum, low2019hamiltonian}.
For the nonadiabatic Hamiltonian in Eq.~\eqref{eq:marcus_hamiltonian}, the block encoding must represent the full $2^n \times 2^n$ potential operators, which are diagonal but dense in the computational basis.
The oracle construction overhead therefore scales with the number of distinct diagonal elements --- $\mathcal{O}(2^n)$ for a general potential on the binary-encoded grid --- rather than with any favorable sparsity structure \cite{babbush2019quantum, martyn2023efficient}.
At $n = 8$ qubits, this translates to block-encoding a two-surface system with $2 \times 256 = 512$ potential values plus the kinetic term, requiring ancilla overhead for the linear combination of unitaries (LCU) decomposition that substantially exceeds the $\mathcal{O}(n^2)$ two-qubit gates needed for explicit Trotterization.
More critically, this oracle requires SWAP routing for all-to-all connectivity, compounding the overhead for hardware with limited native interactions. 

The variational approach we present does not require oracle construction or post-selection. 
The compressed circuits achieve $\mathcal{O}(n)$ depth scaling for quadratic operators, compared to $\mathcal{O}(n^2)$ for explicit Trotterization. 
However, our method requires classical optimization to determine circuit parameters, with optimization cost scaling exponentially for classical simulation of the QAQC protocol (manageable for $n \leq 5$) or polynomially when using direct Hilbert-Schmidt minimization.
For reasonable register sizes ($n\leq 12$), the QAQC method may not be necessary for approximating dynamics on a reaction coordinate.
The emulations of QAQC optimization therefore serve as a proof-of-principle for optimization of all-to-all connective operators.

Once optimized for timestep $\tau$, evolution for $N\tau$ requires no additional optimization, only rescaling of diagonal parameters. 
This enables simulation times far exceeding the coherence time of current hardware, provided the initial optimization achieves sufficient accuracy.
While nonadiabatic dynamics is not fast-forwardable, dynamics in adiabatic regimes may retain sufficient sparsity for accelerated dynamics \cite{atia2017fast}.

\paragraph{Limitations and Sources of Error --}

Several factors limit the accuracy of our compressed simulations.
The $l$-local truncation introduces systematic error by construction, controlled by some threshold $C_{\text{LHST}} < 0.01$. 
This error propagates through Trotter steps, accumulating coherently as $\mathcal{O}(N \cdot C_{\text{LHST}})$ for $N$ steps.
For the rate coefficient calculations presented here, we show that even small truncation error leads to a large deviation in observables, leaving simulation highly sensitive to approximation.

For quantum hardware simulations, gate errors, decoherence, and readout errors degrade fidelity on physical hardware. 
Our fast-forwarding demonstration on $ibm\_brisbane$ showed quick fidelity decay from 0.99 at initialization, but leveling off over longer times, suggesting a possible implementation in adiabatic systems or prior to subjecting the wavefunction to nonadiabatic influences.
Error mitigation techniques such as zero-noise extrapolation could extend the useful simulation depth \cite{temme2017error, kandala2019error}, but significant progress needs to be made before meaningful nonadiabatic dynamics simulations can be applied.
The QAQC cost function may contain local minima, particularly for large $n$ and $l$. 
We verified convergence to global minima by comparing optimized parameters to analytical expressions for all tested configurations, giving the variational parameters of $A^\dagger$ no prior knowledge of $U$. 
For production use, having an unknown global minimum poses practical difficulty, eased by informed initial guesses based on analytical structure.

A related concern for scalability is whether the QAQC cost landscape exhibits barren plateaus at larger $n$ \cite{mcclean2018barren}.
For generic hardware-efficient ansatzes, gradients vanish exponentially with system size, potentially rendering optimization intractable.
However, several features of our approach may mitigate this issue.
First, the target operators are quadratic functions on the binary-encoded grid, which we have shown reduce to two effective variational parameters under $l$-local truncation, regardless of $n$.
This low effective dimensionality constrains the optimization to a structured subspace rather than the full $\mathcal{O}(n^2)$-dimensional parameter space.
Second, the Walsh-based diagonal ansatz is directly motivated by the algebraic structure of the target, in contrast to unstructured ansatzes where barren plateaus are most severe.
Nevertheless, since our classical emulations extend only to $n = 8$, we cannot rule out gradient concentration at larger register sizes, and a systematic study of trainability for this ansatz class remains an important direction for future work.

\section{\label{sec:Conclusions} Conclusions}

We developed a variational method to compress Trotterized quadratic operators for quantum simulation of nonadiabatic molecular dynamics. 
The method employs a VFF ansatz with $l$-local truncation of the Walsh series expansion, optimized using the QAQC protocol. 
The compressed ansatz exactly represents any quadratic operator when $l=n$ for linear qubit topology or $l = \lceil (n+1)/2 \rceil$ for ring topology, achieving $C_{\text{LHST}} = 0$ up to global phase.
Reducing $l$ below these limits introduces bounded approximation error.
Remaining variational parameters compensate during optimization, converging to effective values determined by the polynomial coefficients. 
This reduces the parameter space from $\mathcal{O}(n^2)$ to $2$ for quadratic polynomials.
We see Marcus model rate coefficients computed using truncated ansatzes ($l=4$ for kinetic, $l=6$ for potential terms) retain behavior but deviate from theoretical predictions.
This demonstrates that physically relevant observables are preserved, but are sensitive to controlled circuit approximation.
Recent wavepacket initializations on $ibm\_torino$ achieved fidelities of $0.933 \pm 0.000413$ for $n = 6$ qubits, showing improved error mitigation for larger register operations, but with statistically lower fidelities for $n=3$ (Appendix \ref{app:VQE_init}).
Fast-forwarded dynamics on $ibm\_brisbane$ maintained fidelity above 0.80 through 10 timesteps, validating the practical utility of the compressed ansatz structure.
Explicit gate counts confirm that $l$-local compression reduces circuit depth from $\mathcal{O}(n^2)$ to $\mathcal{O}(n\times l)$, with the locality parameter $l$ determined by the desired approximation accuracy.

Several extensions of this work merit investigation.
The Walsh expansion extends to degree $k > 2$ by including multi-qubit terms ($ZZZ$, $ZZZZ$, etc.). This would enable compression of anharmonic potentials relevant to molecular dynamics beyond the harmonic approximation, including Morse potentials and polynomial fits to ab initio potential energy surfaces \cite{braams2009permutationally, jiang2016potential}.
Our current method compresses operators that are already diagonal in the position basis, for the sole purpose of knowing the global minimum to the cost function. 
Extending to dense operators such as momentum-dependent potentials or non-local couplings would require optimizing the eigenvector unitary $W(\vec{\theta})$ jointly with the diagonal term, increasing the parameter space but potentially enabling compression of entire Trotter steps.
Finally, as quantum hardware transitions toward fault tolerance, comparing variational compression of adiabatic terms to QSP-based methods at equivalent logical error rates will clarify the optimal approach for different problem classes \cite{yoder2025tour, shaw2025lowering}. 
Error correction suggests a need to optimize while minimizing the number of non-transversal gates for fault-tolerant architectures, opening new areas of optimization for quantum operations applied in parallel.

\section{\label{sec:DataAvailability} Data Availability}

Scripts made for this research are provided in a public Github repository \cite{courtney2025vffcompression}.
Data associated with IBM quantum hardware runs are also provided.

\section{\label{sec:Acknowledgements} Acknowledgments}

This work was partially funded by the NSF Research Traineeship Program (NSF grant no. 2152159). We thank Michael Geller for valuable conversation regarding variational optimization.

\appendix

\section{Nonadiabatic Dynamics}
\label{app:nonadiabatic_dynamics}
    The explicit Trotterization follows the explicit expansion of an exponentiated quadratic from \cite{benenti2008quantum}.
    Using the harmonic potential operator $V$ as an example 
    \begin{equation}\label{eq:BenentiDerivation}
        \begin{split}
            e^{-iV\tau}\ket{\psi} & = e^{-i\tau\frac{m\omega^2}{2}(x-x_0)^2 -i\tau\delta}\ket{\psi} \\
            & = e^{-i\tau\frac{m\omega^2}{2}(x^2 - 2xx_0 + x_0^2) -i\tau\delta}\ket{\psi} \\
            & = e^{-i\tau\frac{m\omega^2}{2}x^2}e^{-i\tau\frac{m\omega^2}{2}(-2xx_0)} \\
            & \times e^{-i\tau\frac{m\omega^2}{2}x_0^2 -i\tau\delta}\ket{\psi} \\
            & = e^{-i\tau\frac{m\omega^2\Delta^2}{2}\sum_{\ell=0}^{n-1}\sum_{j=0}^{n-1}2^jk_j2^\ell k_\ell}\\
            & \times e^{-i\tau\frac{m\omega^2}{2}(-2 x_0\Delta\sum_{j=0}^{n-1}2^jk_j)} \\ &\times e^{-i\tau(\frac{m\omega^2}{2}x_0^2 + \delta)}\ket{\psi} \\
            & = e^{2i\tau\eta \Delta^2 \sum_{\ell=0}^{n-1}\sum_{j=0}^{n-1}2^jk_j2^\ell k_\ell} \\
            & \times e^{-i\tau\eta x_0 \Delta\sum_{j=0}^{n-1}2^jk_j}\times e^{-i\tau(\eta x_0^2 + \delta)}\ket{\psi}
        \end{split}
    \end{equation}
    for $x$ being defined by a binary basis $\ket{k_{n-1}\ldots k_1 k_0}$, with $2^n$ discrete positions spanning $x=0$ to $x=L$.
    Here, we define $\eta = m\omega^2 /2$, $\delta = \Delta G$ as the energy difference between quadratic potentials, $x_0$ as the center of the quadratic curve, and $\Delta = L/2^n$ (Figure \ref{fig:explicit_quadratic}).

\begin{figure*}
    \centerline{
    \includegraphics[width=80mm,scale=0.5]{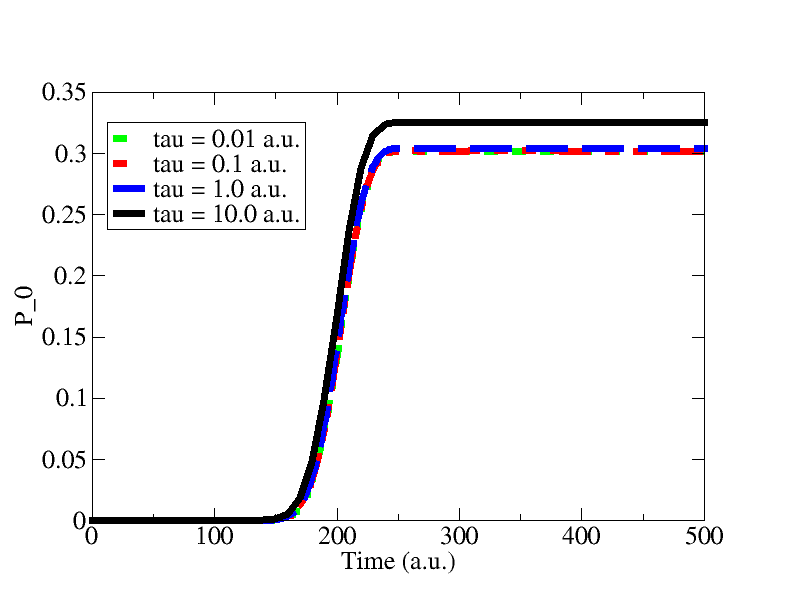}
    \includegraphics[width=80mm,scale=0.5]{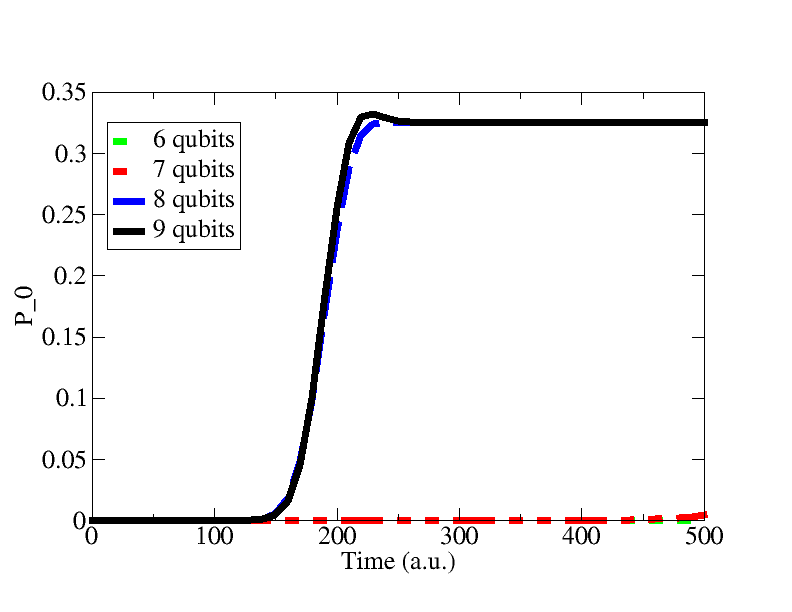}}
    \caption{\label{fig:Adiabatic_varytimestep} 
    Adiabatic state transition in a Marcus-model type system with $p_0 = -30$, $x_0 = 4$. 
    Results are consistent with \cite{ollitrault2020nonadiabatic}, confirming the efficacy of an alternative coupling scheme for varying timesteps (left) and register size (right). 
    The Nyquist limit $\mathcal{N}$ is doubled for every additional qubit in the position register.}
    \end{figure*}

    With the kinetic term operating similarly, we can define our box size as $L=20$, being sufficiently large to contain wavepacket dynamics without aliasing (wrapping due to periodic boundary) in both position and momentum space.
    The momentum ``box'' are given by the Nyquist limits. 
    Wavepackets with greater momenta require sufficient resolution in momentum space for dynamics, evidenced in Figure \ref{fig:Adiabatic_varytimestep} for $p_0 = -30$ and $\Delta G = 0$. 
    Potentials are given according to the relationships $V_0 = A_1(R-L/2 -A_0)^2$ and $V_1 = A_1(R-L/2 +A_0)^2$. 
    These parameters are defined as $A_0 = 1.5$ and $A_1 = 0.015$.

\section{The Global Minimum}
\label{app:global_minimum}
These operations act as phase and controlled phase gates (equivalent to $R_z$ gates) diagonal to the qubit basis. 

$R_z$ gates act as 
\begin{equation}
    R_z(\alpha) \otimes \mathbb{1} = \ket{0}\bra{0}\otimes \mathbb{1} + e^{i\alpha}\ket{1}\bra{1}\otimes \mathbb{1}
\end{equation}
We see that $R_z$ gates therefore commute with other $R_z$ gates
\begin{equation}
    [R_z(\alpha)\otimes\mathbb{1}, R_z(\beta)\otimes\mathbb{1}] = 0
\end{equation}
Likewise, controlled-$R_z$ gates ($CR_z$) act as
\begin{equation}
    CR_z(\alpha) \otimes\mathbb 1 = \ket{0}\bra{0}\otimes\mathbb{1}\otimes\mathbb{1} + \ket{1}\bra{1}\otimes R_z(\alpha) \otimes \mathbb{1}
\end{equation}
Two $CR_z$ gates that act on different tensor spaces, and therefore commute
\begin{equation}
    [CR_z(\alpha) \otimes\mathbb 1, CR_z(\beta) \otimes\mathbb 1] = 0
\end{equation}
So, the quadratic term from Equation \ref{eq:BenentiDerivation} can be reformed as
\begin{equation}
    \begin{split}
        e^{-i\tau\frac{m\omega^2}{2}x^2} & = e^{-i\tau\phi_3 \left(\sum_j^{n-1}2^jk_j\right)^2} \\
        & = e^{-i\tau\phi_3 \left(\sum_j^{n-1}2^{2j}k_j^2 + 2\sum_{j<\ell}2^{j+\ell}k_jk_\ell\right)} \\
        & = e^{-i\tau\phi_3 \left(\sum_j^{n-1}2^{2j}k_j\right)} \\& \times e^{-i\tau\phi_3\left(2\sum_{j<\ell}2^{j+\ell}k_jk_\ell\right)} \\
    \end{split}
\end{equation}
Combining with the first order term, we get 
\begin{equation}
\begin{split}
    & e^{-i\tau\phi_3 \sum_j^{n-1}2^{2j}k_j}\times e^{-i\tau\phi_3\left(2\sum_{j<\ell}2^{j+\ell}k_jk_\ell\right)} \\
    &\times e^{-i\tau\phi_2\sum_j^{n-1}2^jk_j} \\
    & = e^{-i\tau\phi_3\left(2\sum_{j<\ell}2^{j+\ell}k_jk_\ell\right)}\times e^{-i\tau\phi_3 \sum_j^{n-1}2^{2j}k_j}\\
    &\times e^{-i\tau\phi_2\sum_j^{n-1}2^jk_j} \\
    & = e^{-i\tau\phi_3\left(2\sum_{j<\ell}2^{j+\ell}k_jk_\ell\right)}\\
    &\times e^{-i\tau(\phi_3 \sum_j^{n-1}2^{2j}k_j + \phi_2\sum_j^{n-1}2^jk_j)}
\end{split}
\end{equation}

For clarity, $\ell$ is not directly related to $l-$locality and is merely an index.
This expression shows that the compressed ansatz with all terms in the second-order Walsh expansion is equivalent to the explicit ordered operations from \cite{benenti2008quantum}. 
With an $l$-local truncation of the diagonal, parameters within the truncation will ``compensate'' during the optimization such that the first order terms ($R_z$ sublayer) will depend on a single variational parameter, and the second order terms ($ZZ$ sublayers) will depend on a second variational parameter, both related to the parameterization of the quadratic ($\phi_2, \phi_3$). 
The compressed circuit (lower, Figure \ref{fig:explicit_quadratic}) uses phase-dependent controlled-$R_z$ gates. 
As of this writing, it is advantageous to optimize the variational circuit using the native gate set on your desired hardware, even with equivalent Hamiltonian expressivity.
Transpilation does not guarantee the most efficient conversion between, say, a $CR_z$ gate in Qiskit and a parity-based $ZZ(\theta)$ gate applied on a trapped-ion quantum computer. 

\section{Log-Size Comparison Circuit}
\label{app:log-size_comparator}
Comparisons are made by storing the truth value of ``Is the wavepacket greater/less than (number)?''
The (number) is given by the qubit basis, used as binary states defining $2^n$ discrete positions ($\ket{0\ldots0}$ is state $0$, $\ket{0\ldots1}$ is state $1$, etc.). 
The algorithm uses logical AND and OR operations via Toffoli gates.
To test if a quantum state has a value greater than these values, as thresholds logical comparisons can be made.
In the following comparisons, $q_0$ is the $low$ $bit$, as $q = q_{n-1} \cdots q_0$.
\begin{equation}
    \begin{split}
        \ket{\Psi} & > \ket{000} \text{ if } q_2 = 1 \text{ OR } q_1 = 1 \text{ OR } q_0 = 1 \\ 
        \ket{\Psi} & > \ket{001} \text{ if } q_2 = 1 \text{ OR } q_1 = 1\\
        \ket{\Psi} & > \ket{010} \text{ if } q_2 = 1 \text{ OR } (q_1 = 1\text{ AND } q_0 = 1) \\
        \ket{\Psi} & > \ket{011} \text{ if } q_2 = 1 \\
        \ket{\Psi} & > \ket{100} \text{ if } q_2 = 1 \text{ AND } (q_1 = 1 \text{ OR } q_0 = 1) \\
        \ket{\Psi} & > \ket{101} \text{ if } q_2 = 1\text{ AND } q_1 = 1 \\
        \ket{\Psi} & > \ket{110} \text{ if } q_2 = 1\text{ AND } q_1 = 1 \text{ AND } q_0 = 1
    \end{split}
    \label{eq:logic}
\end{equation}

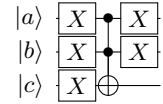
\begin{figure}
\centerline{
\Qcircuit @C=0.1em @R=0.1em {
\lstick{\ket{a}} & \gate{X} & \ctrl{2} & \gate{X} & \qw \\
\lstick{\ket{b}} & \gate{X} & \ctrl{1} & \gate{X} & \qw \\
\lstick{\ket{c}} & \gate{X} & \targ & \qw & \qw
}}
\caption{\label{fig:LogicalOR} A logical OR operation.
If $\ket{a}$ or $\ket{b}$ holds the state 1, then $\ket{c}$ will be in state 1 after the circuit. 
Otherwise, the Toffoli gate will ``turn off'' $\ket{c}$ so it is 0.}
\end{figure}

Quantum circuits can be constructed to test each of these using a logical OR, shown in Figure \ref{fig:LogicalOR}.

\begin{figure*}
 \includegraphics[width=100mm]{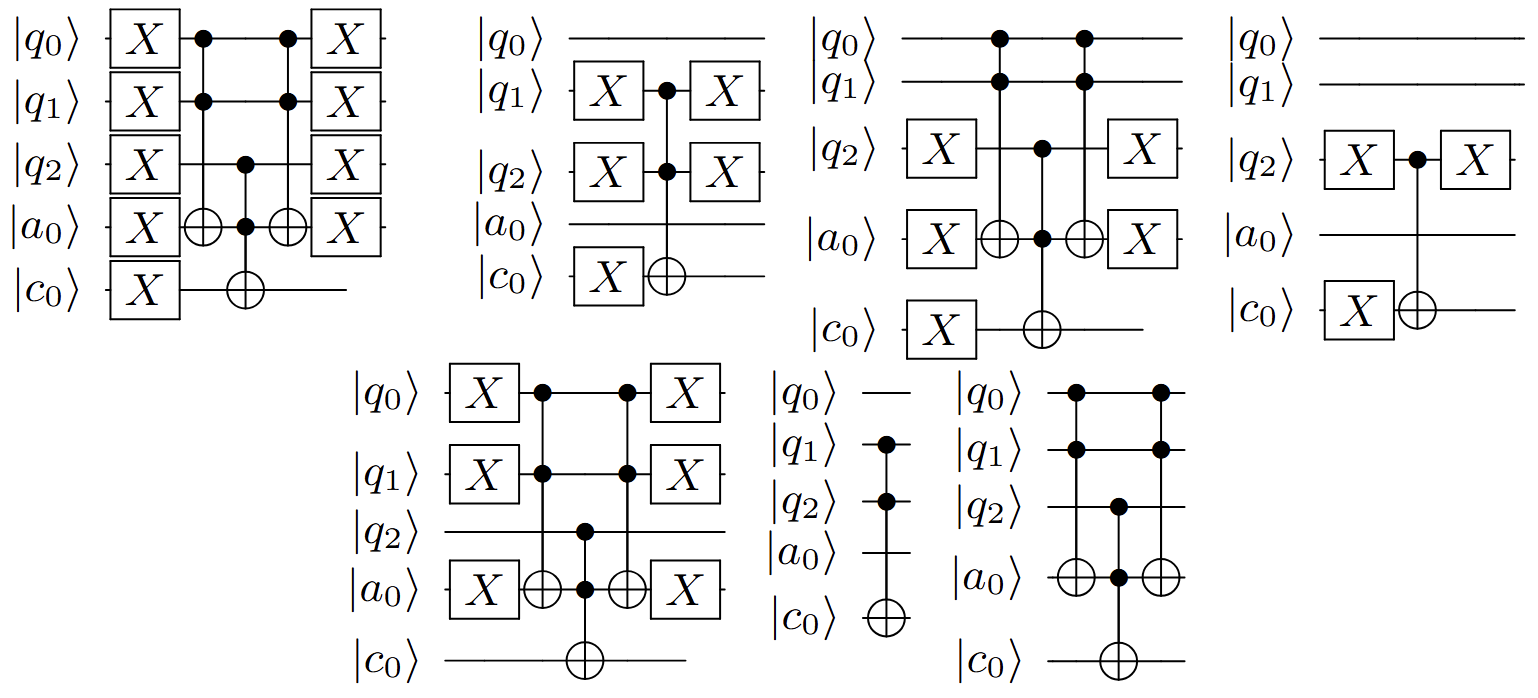}
\caption{\label{fig:3QubitComparisons} Comparison circuits to test if a 3-qubit state is ``greater than'' a given integer value from 0 (state $\ket{000}$, top left) to 6 (state $\ket{110}$, bottom right). 
A test for ``greater than'' state $\ket{111}$ is meaningless in this basis, though $<$, $=$, $\leq$, and $\geq$ operands can be constructed similarly. 
For a single comparison the minimum number of qubits that need to be checked is always one (see $>3$, top right), occurring only once in a set of $\mathcal{N}$ basis states, and the maximum number to be checked is always $n$, occurring as half of all basis states.
The number of comparisons that need to check $m$ qubits is $2^{m-1}$ for $m \leq n$.}
\end{figure*}

To temporarily store the result of a logical operation, an ancilla qubit $\ket{a_0}$ is required.
Inverse operations can be performed on the ancilla to return it to the $\ket{0}$ state for future use, though reinitialization (aka a ``reset'' operation) may be required in later practice to prevent uncorrectable state-mixing between ancillary, comparison, and position registers.
One comparator qubit $\ket{c_0}$ is needed to store the resulting truth value of each value comparison. For a state represented by three qubits, therefore, a total register size of five is needed. 
The test for $>2$ is given in Stamatopolous \cite{stamatopoulos2020option}. 
The seven operations in Equation \eqref{eq:logic} therefore translate to quantum circuits given Figure \ref{fig:3QubitComparisons}. 

To test if a state is less than a threshold value, the logic associated to the one's complement to a binary number can be applied.
To clarify, logical conditions defining $>$ for a state $\ket{\Psi}$ are identical to the one's complement of each logical condition for defining $<$ for a state $\ket{\Psi'}$, where $\ket{\Psi'}$ is the one's complement of $\ket{\Psi}$. 

Effectively, a single OR or AND statement requires one Toffoli gate.
With three qubits, a maximum of three Toffoli gates are needed, (1) to write the truth value between two qubits to an ancilla, one to write the truth value between the third qubit and the ancilla to the comparator, and one to un-write the state of the ancilla.
The final ``un-write'' operation can be removed if the comparison will be un-written later.
The methods shown here clearly do not consider state mixing between registers as an inevitable reality in hardware simulations, but are constructed to avoid the use of a ``reset'' gate. 

The maximum number of AND or OR statements that must be applied to perform the $<$ or $>$ operation will check the state of every qubit. 
The maximum number of qubits that can be checked at once is two (CCNOT gate aka Toffoli gate). 
With these constraints, the miminum number of ancillary qubits scales as $\mathcal{O}(\log_2 n)$. Depth scaling for this size scaling is $2n^{1.5}/3$ $+\mathcal{O}(1)$. With additional ancillary qubits up to $n_{anc} \propto \mathcal{O}(n)$, it is easy to find that depth scaling is reduced to $\mathcal{O}(\log_2 n)$.

\section{Coupling Function}

\begin{figure*}
\centerline{
\Qcircuit @C=0.1em @R=0.5em {
\lstick{\ket{\text{q}_0}}& \qw & \ctrl{8} & \qw & \qw & \qw & \ctrl{8}&\qw & \qw & \qw & \qw & \qw & \qw &\qw & \qw & \qw & \qw & \qw & \qw & \ctrl{8} & \qw & \qw & \qw & \ctrl{8} & \targ & \ctrl{8} & \qw & \qw & \qw & \ctrl{8} & \qw & \qw & \qw & \qw& \qw & \qw & \qw\\
\lstick{\ket{\text{q}_1}}& \qw & \ctrl{7} & \qw & \qw & \qw & \ctrl{7}&\qw & \qw & \qw & \qw & \qw & \qw &\qw & \qw & \qw & \qw & \qw & \qw & \ctrl{7} & \qw & \qw & \qw & \ctrl{7} & \targ & \ctrl{7} & \qw & \qw & \qw & \ctrl{7} & \qw & \qw & \qw & \qw& \qw & \qw& \qw\\
\lstick{\ket{\text{q}_2}}& \qw & \qw & \ctrl{7} & \qw & \ctrl{7} & \qw & \qw & \qw & \qw & \qw & \qw & \qw & \qw &\qw & \qw & \qw & \qw & \qw & \qw & \ctrl{7} & \qw & \ctrl{7} & \qw & \targ & \qw& \ctrl{7} & \qw & \ctrl{7} & \qw  & \qw & \qw & \qw & \qw& \qw & \qw\\
\lstick{\ket{\text{q}_3}}& \qw & \qw & \ctrl{6} & \qw & \ctrl{6} & \qw & \qw & \qw & \qw & \qw & \qw & \qw & \qw &\qw & \qw & \qw & \qw & \qw & \qw & \ctrl{6} & \qw & \ctrl{6} & \qw & \targ & \qw & \ctrl{6} & \qw & \ctrl{6} & \qw & \qw & \qw & \qw & \qw& \qw & \qw \\
\lstick{\ket{\text{q}_4}}& \qw & \qw & \qw & \qw & \qw & \qw & \ctrl{4} & \qw & \qw & \ctrl{4}& \qw & \qw & \qw & \qw & \ctrl{4} & \qw & \qw& \ctrl{4}& \qw & \qw & \qw & \qw & \qw & \targ & \qw & \qw & \qw & \qw & \qw & \ctrl{4} & \qw & \ctrl{4} & \qw & \qw & \qw & \qw\\
\lstick{\ket{\text{q}_5}}&\qw & \qw & \qw & \qw & \qw & \qw & \ctrl{3} & \qw & \qw & \ctrl{3}& \qw & \qw & \qw & \qw & \ctrl{3} & \qw & \qw& \ctrl{3}& \qw & \qw & \qw & \qw & \qw  & \targ & \qw & \qw & \qw & \qw & \qw & \ctrl{3} & \qw & \ctrl{3} & \qw & \qw & \qw & \qw\\
\lstick{\ket{\text{q}_6}}& \qw & \qw & \qw & \qw & \qw & \qw & \qw & \qw & \qw & \qw & \ctrl{2} & \qw & \qw & \ctrl{2}& \qw & \qw & \qw & \qw & \qw & \qw & \qw & \qw & \qw & \targ & \qw & \qw & \qw & \qw & \qw & \qw & \ctrl{3} & \qw & \qw & \qw & \qw & \qw\\
\lstick{\ket{\text{q}_7}}&\targ& \qw & \qw & \qw & \qw & \qw & \qw & \qw & \qw & \qw & \qw & \ctrl{4} & \ctrl{5} & \qw& \qw & \qw & \qw & \qw & \qw & \qw & \qw & \qw & \qw & \targ & \qw & \qw & \qw & \qw & \qw & \qw & \qw & \qw & \qw & \qw &\ctrl{6} & \qw\\
\lstick{\ket{0}}& \qw & \targ & \qw & \ctrl{2} & \qw & \targ & \targ & \targ & \ctrl{1} & \targ & \targ & \ctrl{3} & \qw & \targ & \targ & \ctrl{1} & \targ & \targ & \targ & \qw & \ctrl{2} & \qw & \targ & \qw & \targ & \qw & \ctrl{2} & \qw & \targ & \targ & \ctrl{1} & \targ & \targ & \qw & \ctrl{3} & \qw \\
\lstick{\ket{0}}& \qw & \qw & \targ & \ctrl{1} & \targ & \qw & \qw & \qw & \targ & \qw & \ctrl{-1} & \qw & \qw & \ctrl{-1} & \qw & \targ & \qw & \qw & \qw & \targ & \ctrl{1} & \targ & \qw & \qw & \qw & \targ & \ctrl{1} & \targ & \qw & \qw & \targ & \qw & \ctrl{-1} & \qw & \qw & \qw\\
\lstick{\ket{0}}& \qw & \qw & \qw & \targ & \qw & \qw & \qw & \qw & \ctrl{-1} & \qw & \qw & \qw & \qw & \qw & \qw & \ctrl{-1} & \qw & \qw & \qw & \qw & \targ & \qw & \qw & \qw & \qw & \qw & \targ & \qw & \qw & \qw & \qw & \qw & \ctrl{-2} & \qw & \qw & \qw\\
\lstick{\ket{c_0}}& \qw & \qw & \qw & \qw & \qw & \qw & \qw & \qw & \qw & \qw & \qw & \targ & \qw & \qw & \qw & \qw & \qw & \qw & \qw& \qw & \qw & \qw & \qw & \qw & \qw & \qw & \qw & \qw & \qw & \qw & \qw & \qw & \qw & \qw & \qw & \qw \\
\lstick{\ket{c_1}}&\targ & \qw & \qw & \qw & \qw & \qw & \qw & \qw & \qw & \qw & \qw & \qw & \targ & \qw & \qw & \qw & \qw & \qw & \qw & \qw & \qw & \qw & \qw & \qw & \qw & \qw & \qw & \qw & \qw & \qw & \qw & \qw & \qw & \qw & \qw & \qw \\
\lstick{\ket{c_2}}&\targ& \qw & \qw & \qw & \qw & \qw & \qw & \qw & \qw & \qw & \qw & \qw & \qw & \qw & \qw & \qw & \qw & \qw & \qw & \qw & \qw & \qw & \qw & \qw & \qw & \qw & \qw & \qw & \qw & \qw & \qw & \qw & \qw & \qw & \targ & \qw
}
}
\caption{\label{fig:Greater-Than} Comparison circuits to test if an 8-qubit state is ``less than'' basis states 127, 128, and 129.}
\end{figure*}
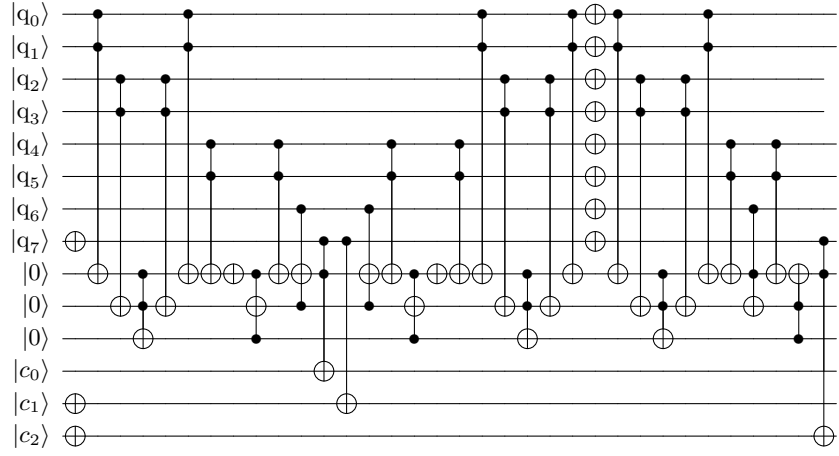

Ollitrault et al. (2020) used an 8-qubit register to perform a simulated Trotter-Suzuki evolution of a particle in a coupled quadratic potential \cite{ollitrault2020nonadiabatic}. 
The coupling was varied, one of which used three breakpoints. 
In Figure \ref{fig:Greater-Than} is a circuit used in the current work to determine if the eight-qubit quantum state $\ket{\Psi}$ is less than three values, being 127, 128, and 129. 'Less than' is chosen since the wavepacket is propagating from right to left in the simulation.
For brevity, the ancilla are not freed in the given circuit, but are freed after the coupling has been applied. 
The $2^n =256$ discrete positions represent a space of $L = 20$. 
Results for adiabatic state transfer ($\Delta G = 0$) between $V_0$ and $V_1$ is given in Figure \ref{fig:Adiabatic_varytimestep} for varying timesteps $\tau$ and numbers of qubits $n$. 
Characterization of the system is given in the following section. For direct comparison with \cite{ollitrault2020nonadiabatic}, a step function with breakpoints of 127 and 129 was used for nonadiabatic molecular dynamics simulations, given by the same circuit as Figure \ref{fig:Greater-Than} without the CNOT to qubit $c_1$.

After making each comparison, the $objective$ qubit has rotations about the $x$-axis controlled by the position register and the comparison qubits, given by the desired piecewise linear approximation, as 
\begin{equation}
    f(x) = \sum_{i=0}^{P-1} \alpha_i x + \beta_i
\end{equation}

\begin{figure*}\label{fig:Four-Piece}
\centerline{
\Qcircuit @C=0.1em @R=0.1em {
\lstick{\text{position}}& \multigate{4}{x > bp_i\text{?}} & \qw & \ctrl{1} & \ctrl{1} & \qw & \qw & \ctrl{1} & \ctrl{1} & \qw & \qw & \ctrl{1} & \ctrl{1} & \qw & \qw & \ctrl{1} & \multigate{4}{[x > bp_i\text{?}]^{-1}}\\
\lstick{\text{objective}}& \ghost{x > bp_i\text{?}} & \gate{\beta_0} & \gate{\alpha_0} & \gate{-\alpha_0} & \gate{-\beta_0} & \gate{\beta_1} & \gate{\alpha_1} & \gate{-\alpha_1} & \gate{-\beta_1} & \gate{\beta_2} & \gate{\alpha_2} & \gate{-\alpha_2} & \gate{-\beta_2} & \gate{\beta_3} & \gate{\alpha_3} & \ghost{[x > bp_i\text{?}]^{-1}}\\
\lstick{\text{c}_0}& \ghost{x > bp_i\text{?}} & \qw & \qw & \ctrl{-1} &  \ctrl{-1} & \ctrl{-1} & \ctrl{-1} &\qw & \qw & \qw & \qw & \qw & \qw & \qw & \qw & \ghost{[x > bp_i\text{?}]^{-1}}\\
\lstick{\text{c}_1}& \ghost{x > bp_i\text{?}} & \qw & \qw &  \qw & \qw &  \qw & \qw & \ctrl{-2} & \ctrl{-2} & \ctrl{-2} & \ctrl{-2}\qw & \qw & \qw & \qw & \qw & \ghost{[x > bp_i\text{?}]^{-1}}\\
\lstick{\text{c}_2}& \ghost{x > bp_i\text{?}} & \qw & \qw & \qw & \qw & \qw & \qw & \qw & \qw & \qw & \qw & \ctrl{-3} & \ctrl{-3} & \ctrl{-3} & \ctrl{-3} & \ghost{[x > bp_i\text{?}]^{-1}}\\
}
}
\caption{\label{fig:Four-Piece Coupling} Schematic coupling for each piece of the function, conditioned upon the state of the wavepacket and the state of comparator qubits.}
\end{figure*}
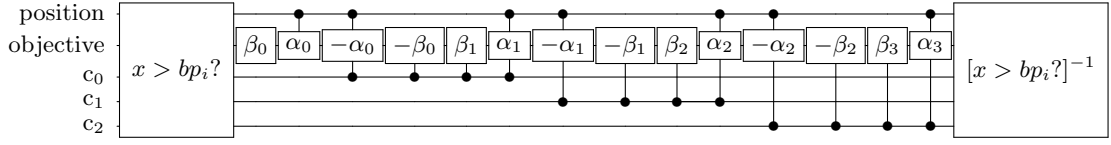

\begin{figure*}
    \centerline{
    \includegraphics[width=60mm,scale=0.5]{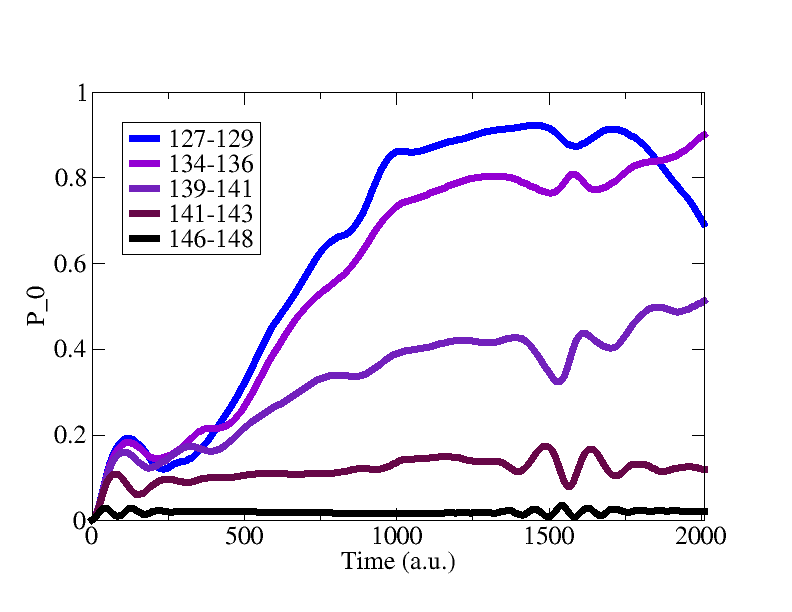}
    \includegraphics[width=60mm,scale=0.5]{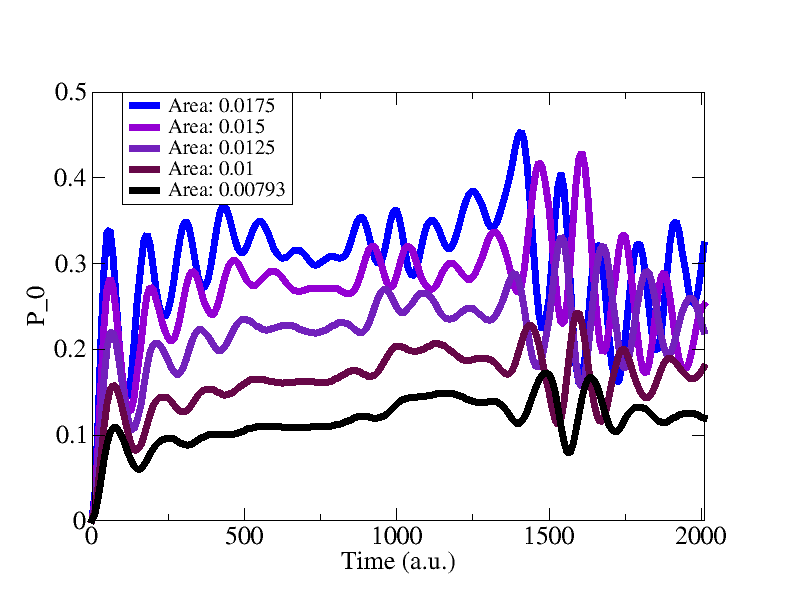}}
    \centerline{
    \includegraphics[width=60mm,scale=0.5]{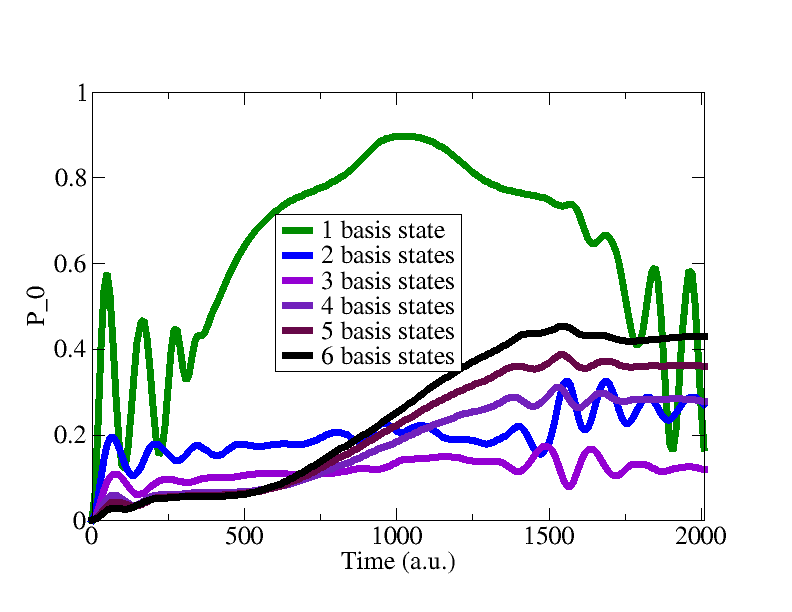}
    \includegraphics[width=60mm,scale=0.5]{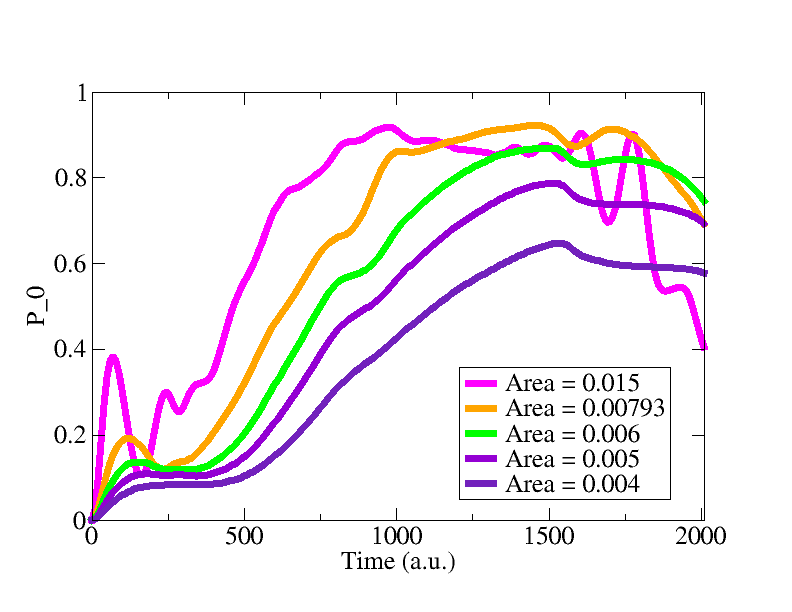}}
    \centerline{
    \includegraphics[width=60mm,scale=0.5]{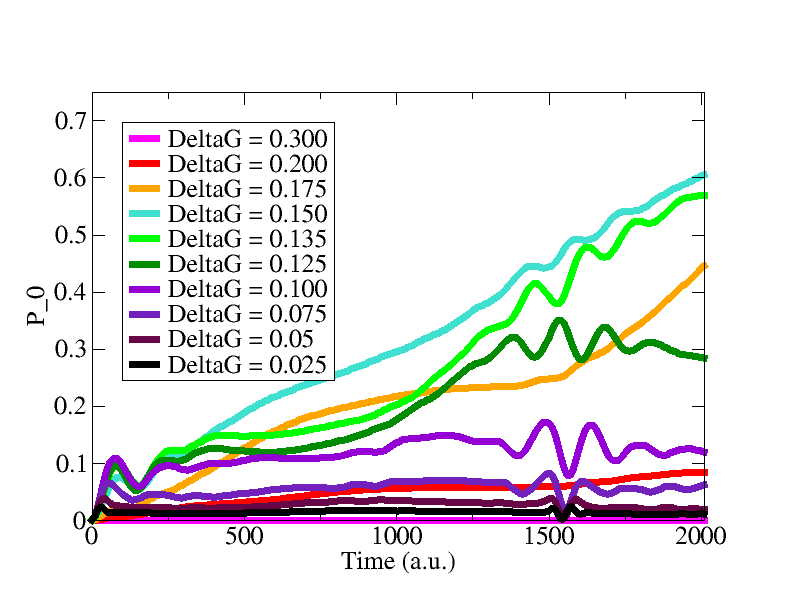}
    \includegraphics[width=60mm,scale=0.5]{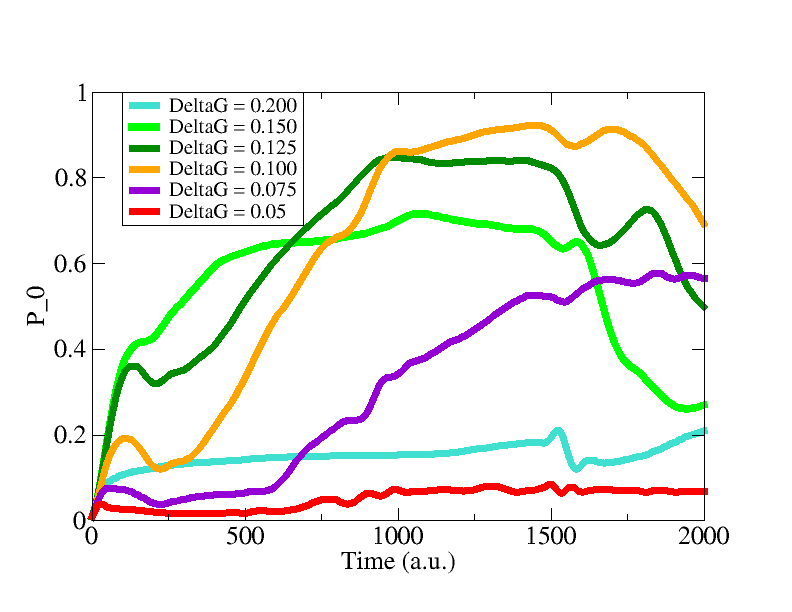}
    }
    \caption{\label{fig:VaryCoupling} 
    Long time dynamics with varying coupling.
    (Top left) Coupling is area preserving with an offset $\Delta G = 0.1$, spanning three basis states, including the midpoint (127-129) and the intersection (134-136).
    (Top right) Coupling centered at the intersection between potential surfaces with varying area with constant offset $\Delta G = 0.1$, spanning three states.
    Area of 0.00793 is the area-preserving result to an approximate Gaussian coupling.
    (Middle left) Area preserving coupling with varying width with constant offset $\Delta G = 0.1$.
    (Middle Right) Coupling centered at the midpoint (basis states 127-129) between potential surfaces with varying area with constant offset $\Delta G = 0.1$, spanning three states. 
    (Bottom Left) Coupling centered at the intersection between potential surfaces for varying energy offsets, in contrast to varying energy offset for coupling between surfaces occurring at the midpoint (bottom right).
    We find closest resemblance to \cite{ollitrault2020nonadiabatic} with a step coupling spanning three basis states, being area-preserving to the exact Gaussian form, and always centered at the midpoint of the qubit basis.}
    \end{figure*}

By the condition of ``if the wavepacket is less than position i,'' we can conditionally apply a linear coupling to the entire space of $2^n$ basis states. 
We can subsequently ``turn off'' this coupling via the truth values of other comparison qubits.
A schematic of this coupling for a $P=4$-piece linear approximation of $f(x)$ is given in Figure \ref{fig:Four-Piece Coupling}.
Varying coupling parameterizations drastically changes the molecular dynamics. 
Such variations are given in Figure \ref{fig:VaryCoupling}.

For all dynamics present in the main text, the coupling was a step function that preserves the area of the Gaussian $Ce^{-\beta (x-a)^2}$ with $C = 0.01$, $\beta = 5.0$, and $a = L/2$ for all values of $\Delta G$.
This step function occupied three basis states (127, 128, 129).
In other words, the coupling did not follow the intersection of the two harmonic potentials, in accordance with results from previous research \cite{ollitrault2020nonadiabatic}. 

In this way, the minimal circuit size for nonadiabatic molecular dynamics that preserves wavepacket structure becomes $8+\log_2(8) + 3 +\log_2(2) = 15$ qubits, with the final qubit being the $objective$ qubit storing state populations of $\kappa = 2$ potential surfaces, giving an improvement from the previously stated minimal circuit size of 18 qubits in \cite{ollitrault2020nonadiabatic}. 

\section{Quantum Assisted Quantum Compiling}
    Cost for the Local Hilbert Schmidt test is given by the sum of entanglement fidelities between two circuits. 
    \begin{equation}
        C_{LHST}(U,V) = 1 - \frac{1}{n}\sum_{j=1}^nF_e^j
    \end{equation}
    
    Bell pairs are initialized between qubits $A_j$ and $B_j$ in circuits $A$ and $B$, where the target unitary $U$ and ansatz $V$ are operating, respectively. 
    The entanglement fidelity for a qubit pair $F_e^{(j)}$ traces out the rest of the quantum registers, as 
    \begin{equation}
    \begin{split}
        F_e^{(j)} &:= \text{Tr}\Big( \ket{\Phi^+}\bra{\Phi^+}_{A_j B_j} (\mathcal{E}_j \otimes \mathcal{I}_{B_j})\\&(\ket{\Phi^+}\bra{\Phi^+}_{A_jB_j})\Big)
    \end{split}
    \end{equation}

    $\mathcal{E}_j$ is the quantum channel acting on $A_j$. With $\bar{A}_j$ as all qubits in $A$ except qubit $j$,
    \begin{equation}
        \mathcal{E}_j(\rho_{A_j}) = \text{Tr}_{\bar{A}_j}\left(\rho_{A_j} \otimes \frac{\mathbb{1}_{\bar{A}_j}}{2^{n-1}}VU^\dagger\right).
    \end{equation}

     We find faithfulness conditions and error propagation in \cite{cirstoiu2020variational} and noise resilience in \cite{khatri2019quantum, sharma2020noise}.

     The fully classical emulation of QAQC uses the evaluations of cost and gradients in the main text, and feeds back into subsequent iterations. 
     An updated vector $\vec{\alpha} = (\vec{\theta}, \vec{\gamma})$ is found by gradient descent, with values for the learning rate updated via Adam optimization. 

     Such an optimization dynamically fits the learning rate to the cost magnitude, allowing for a ``patience'' if a cost $C_{LHST}' -C_{LHST} < \epsilon$. 
     Adam offers bias correction and robustness to the gradient descent to leave local minima and not get caught in barren plateaus, events common in a parameter space with effective dimension of $\mathcal{O}(n^2)\mod2\pi$.

    Time required to classically simulate a quantum circuit grows exponentially with each additional qubit. 
    In our implementation, some QAQC optimizations took upwards of 70 core hours to simulate for a $5$-qubit target unitary (distributed between 4 AMD EPYC Milan (3rd gen) processors, capped at 256GB total memory).
    For ease, this was simulated using Qiskit packages in Python.
    Though improvements on our implementation can surely be made, including the use of quantum simulation packages native to lower-level languages, the exponential behavior to simulating quantum systems remains prohibitive.
    So, the Hilbert Schmidt distance $||V^\dagger U||$ was minimized for larger quantum circuits $n=6-8$, using numerical evaluations of cost and gradients with the same Adam gradient descent prescription as with the QAQC emulator.

    \section{Wavepacket Initialization}
    \label{app:VQE_init}
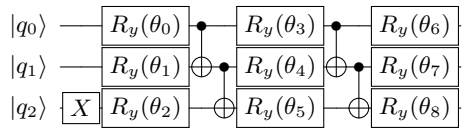
\begin{figure*}
    \centerline{
    \Qcircuit @C=0.1em @R=0.1em {
    \lstick{\ket{q_0}}& \qw & \gate{R_y(\theta_0)} & \ctrl{1} & \qw & \gate{R_y(\theta_3)} & \ctrl{1} & \qw & \gate{R_y(\theta_6)} & \qw\\
    \lstick{\ket{q_1}}& \qw & \gate{R_y(\theta_1)} & \targ & \ctrl{1} & \gate{R_y(\theta_4)} & \targ & \ctrl{1} & \gate{R_y(\theta_7)} & \qw\\
    \lstick{\ket{q_2}}& \gate{X} & \gate{R_y(\theta_2)} & \qw & \targ & \gate{R_y(\theta_5)} & \qw & \targ & \gate{R_y(\theta_8)} & \qw }}
    \caption{\label{fig:UCC ansatz} An $n=3$-qubit UCC ansatz circuit to initialize a Gaussian wavefunction.}
    \end{figure*}

    We follow a UCC ansatz circuit as given in \cite{ortiz2001quantum}.
    This circuit is given in Figure \ref{fig:UCC ansatz} and represents the initialization of a static wavepacket.
    
\begin{figure*}
    \centerline{
    \includegraphics[width=\textwidth,scale=0.5]{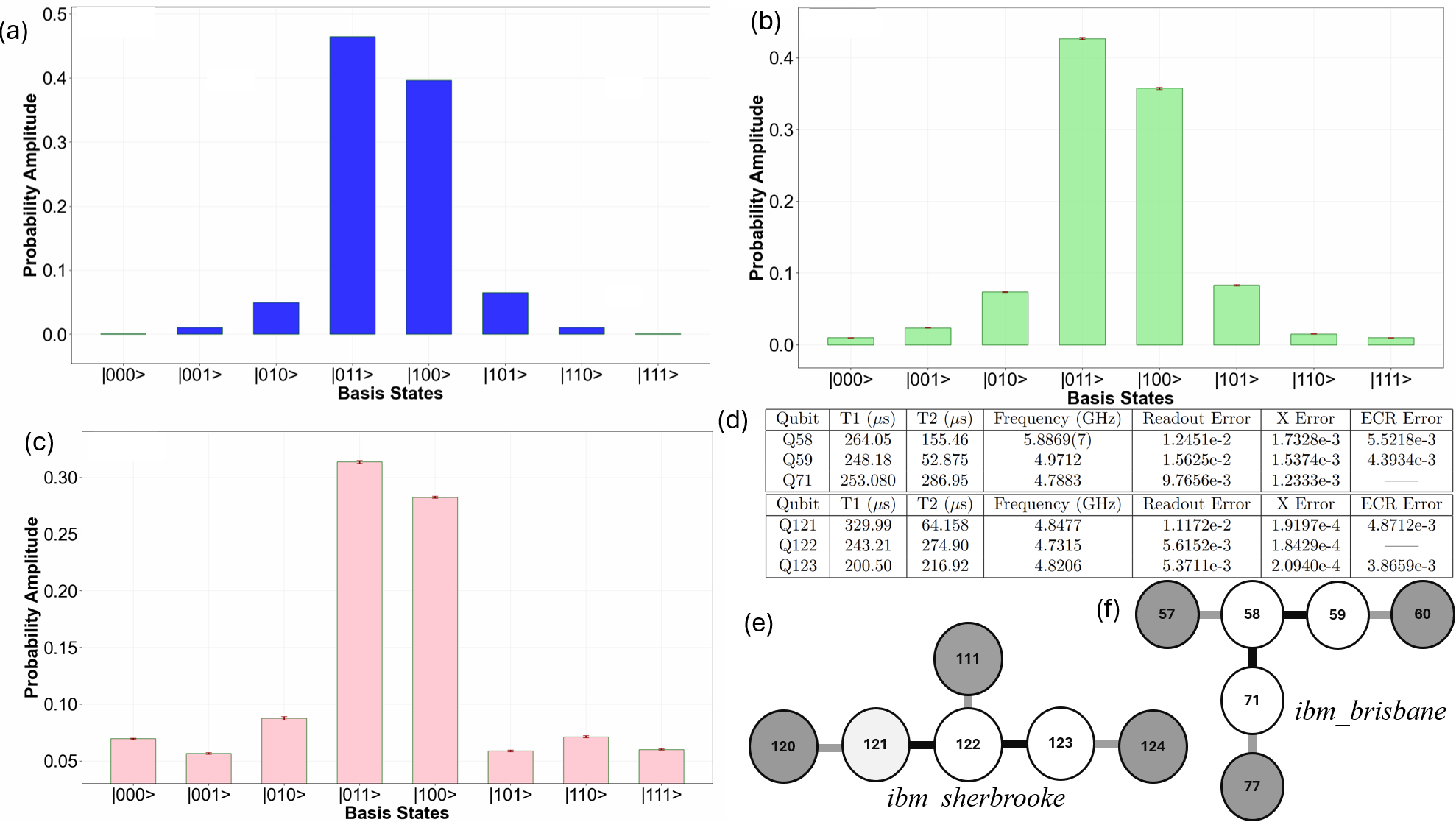}}
    \caption{\label{fig:3q_initialization} 
    Three-qubit initializations using a UCC ansatz circuit.
    Probabilities are given for each basis state using (a) a noise-free simulator, (b) $ibm\_brisbane$ ($F = 0.991 \pm 0.000623$), and (c) $ibm\_sherbrooke$ ($F= 0.912 \pm 0.00129$).
    (d) Qubit connectivities and calibrations for present simulations.}
    \end{figure*}

A classical gradient descent optimization was performed to initialize this wavepacket, as the solution to a harmonic potential $V = a(x-c)^2$, centered in the middle of the qubit basis.
As the beginning to the fast-forwarding ansatz $(t = 0)$, diagnostic measurements were taken on IBM Quantum hardware, being $ibm\_brisbane$ and $ibm\_sherbrooke$. 
Ten independent trials with 1024 shots each were taken on each chip, with results given in Figure \ref{fig:3q_initialization}. 

\begin{figure*}
    \centerline{
    \includegraphics[width=120mm,scale=0.5]{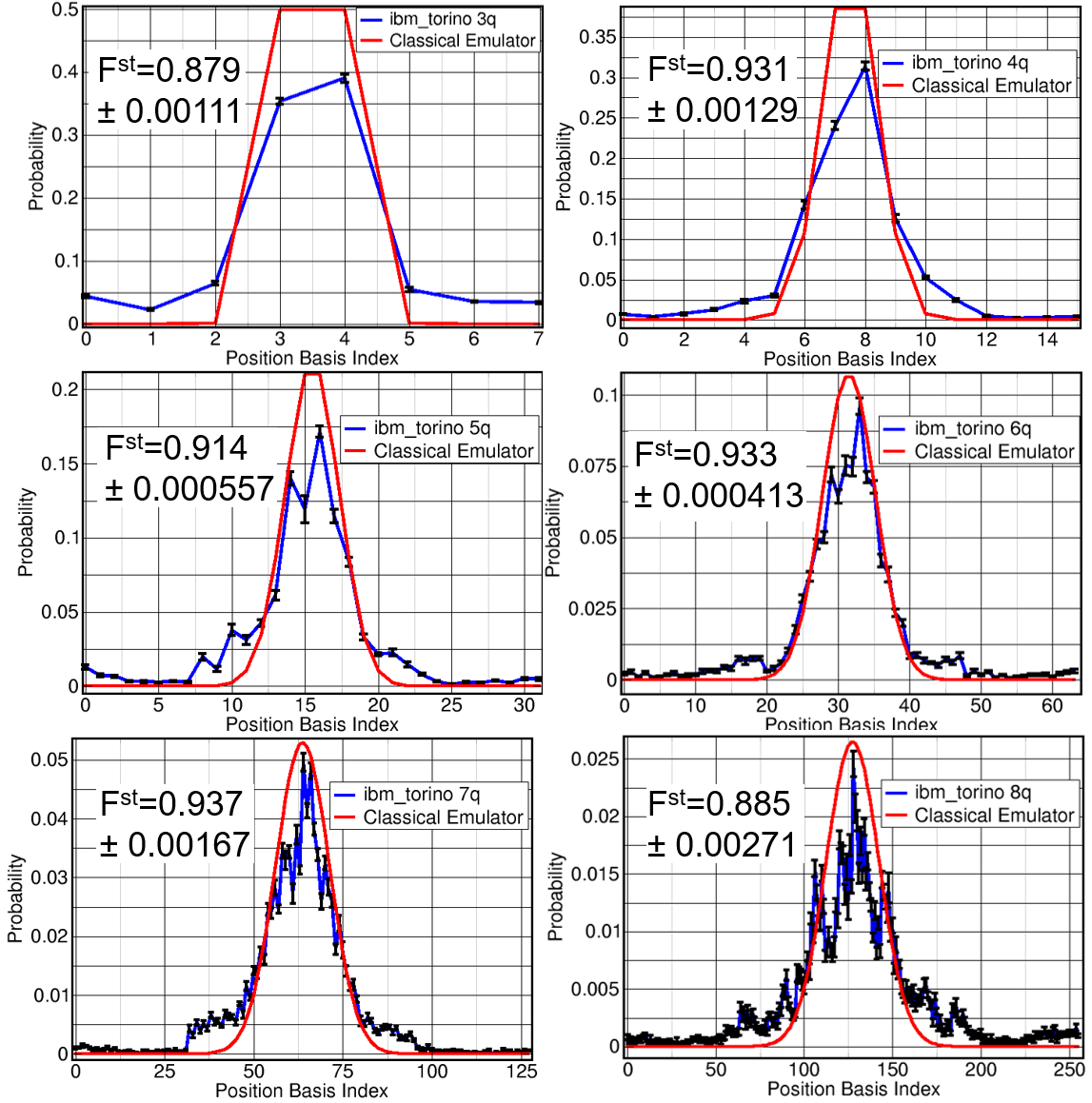}}
    \caption{\label{fig:3q_6q_torino} 
    Wavepacket initializations using a UCC ansatz circuit for $n=3-8$ qubits.
    Probabilities are given for each basis state using average results for $ibm\_torino$ (blue) a classical emulator (red). 
    Error bars for $ibm\_torino$ were taken as the standard error over ten independent runs (job submissions) of 1024 shots each (2048 shots for $n=7-8$).
    State fidelities are shown as insets in each plot.}
    \end{figure*}

    To compare state fidelity between quantum hardware and classical emulation, ten independent trials with 1024 shots each were taken on $ibm\_torino$, with qubit register sizes ranging from $n=3$ to $n=8$ (Figure \ref{fig:3q_6q_torino}). 

\bibliography{NonAdiaRefs}

\end{document}